\documentclass[twocolumn]{autart}    % Enable this line and disable the
\newtheorem{theorem}{Theorem}
\newtheorem{lemma}{Lemma}
\newtheorem{corollary}{Corollary}
\usepackage{color}
\usepackage{cite}
\usepackage{amsmath,amssymb,amsfonts}
\usepackage{graphicx}
\usepackage{textcomp}

\usepackage{graphicx}          % Include this line if your
\usepackage{epstopdf}

\usepackage{mathrsfs}

\begin{document}
\begin{frontmatter}

\title{$\mathscr{H}_2$ Model Reduction for Augmented Model of Linear Non-Markovian Quantum  Systems} % Title, preferably not more

\thanks[footnoteinfo]{This work is supported by the National Natural Science Foundation of China (NSFC) under Grants No. 62273226 and No. 61873162.}

\author[GG1,GG2,GG3]{Guangpu Wu},%\ead{18716032768@sjtu.edu.cn},    %,GG3 Add the
\author[GG1,GG2,GG3]{ Shibei Xue},\ead{shbxue@sjtu.edu.cn}              %,GG3
\author[GG4]{ Guofeng Zhang},
\author[GG5]{ Rebing Wu},
\author[GG6]{ Min Jiang},
\author[GG7]{ Ian R. Petersen}

\address[GG1]{School of Automation and Intelligent Sensing, Shanghai Jiao Tong University, Shanghai 200240, People's Republic of China}
\address[GG2]{Key Laboratory of System Control and Information Processing, Ministry of Education of China, Shanghai 200240}

\address[GG3]{Shanghai Key Laboratory of Perception and Control in Industrial Network Systems, Shanghai, 200240}

%\address[GG3]{Shanghai Engineering Research Center of Intelligent Control and Management, Shanghai 200240, P. R. China}
\address[GG4]{Department of Applied Mathematics, The Hong Kong Polytechnic University, Hung Hom, Kowloon, Hong Kong Special Administrative Region, People's Republic of China}
\address[GG5]{Center for Intelligent and Networked Systems, Department of Automation, Tsinghua University,
Beijing 100084, People's Republic of China}
\address[GG6]{School of Electronics and Information Engineering,
Soochow University, Suzhou 215006, People's Republic of China}
\address[GG7]{School of Engineering,  Australian National University, Canberra, ACT 2601, Australia}
% Please supply

%
%
%\affiliation{$^{1}$Department of Automation, Shanghai Jiao Tong University, Shanghai 200240, P. R. China}
%\affiliation{$^{2}$Key Laboratory of System Control and Information Processing, Ministry of Education of China, Shanghai 200240, P. R. China}
%\affiliation{$^{3}$Shanghai Engineering Research Center of Intelligent Control and Management, Shanghai 200240, P. R. China}
%\affiliation{$^{4}$Department of Mechanical and Automation Engineering, The Chinese University of Hong Kong, Shatin, Hong Kong SAR, P. R. China}
%\affiliation{$^{5}$School of Physics and Electronics, Hunan Normal University, Changsha 410081, P. R. China}

                             % keyword list or with the
                                          % help of the Automatica
                                          % keyword wizard

\begin{abstract}
An augmented system model provides an effective way to model non-Markovian quantum systems, which is useful in filtering and control for this class of systems. However, since a large number of ancillary quantum oscillators representing internal modes of a non-Markovian environment  directly interact with the principal system in these models, the dimension of the augmented system may be very large causing significant computational burden in designing  filters and controllers. In this context, this paper proposes an $\mathscr{H}_2$  model reduction method for the augmented model of linear non-Markovian quantum systems.
We first establish  necessary and sufficient conditions for the physical realizability of the augmented model of linear non-Markovian quantum systems, which are more stringent than those for Markovian quantum systems. However, these physical realizability conditions of augmented system model pose non-convex constrains in the optimization problem of model reduction, which makes the problem different from the corresponding classical model reduction problem.
To solve the problem, we derive  necessary conditions for determining the input matrix in the reduced model, with which a theorem for designing the system matrix of the ancillary system in the reduced system is proved. Building on this, we convert the nonlinear equality constraints into inequality constraints so that a semidefinite programming algorithm can be developed to solve the optimization problem for model reduction.
 A numerical example of a two-mode linear quantum system driven by three internal modes of a non-Markovian environment validates the effectiveness of our method.
\end{abstract}

\end{frontmatter}

\section{Introduction}\label{sec1}

As quantum information processing advances rapidly, quantum control has become a crucial and fundamental technique for achieving efficient and precise manipulation of quantum systems \cite{bg1}. The basis of quantum control lies in the accurate and effective modelling of the dynamics of quantum systems \cite{bg2,bg3}.
 However, in practice, since quantum systems often exhibit extremely high dimensionality, the corresponding computational resource requirements  pose significant bottlenecks to the efficiency of quantum control. Therefore, model reduction has attracted considerable attention in the field of quantum control, whose main task is to preserve the core dynamics of the system of interest while significantly reducing the dimensionality of the model so as to improve computational complexity.

For model reduction of closed quantum systems, Ref. \cite{kumar2014} has proposed  a physically relevant invariant subspace spanned by an initial state and parameterized Hamiltonian, which can describe the major dynamics of finite-dimensional systems. Since actual quantum systems are open, a quantum system inevitably interacts with an environment resulting in decoherence \cite{br}.
Hence, it is  important to consider open quantum systems. For Markovian quantum systems where the energy of a quantum system dissipates to its environment unidirectionally, a series of model reduction methods have been proposed for linear quantum systems.
%Among them, Ref. \cite{re1} has developed a singular-perturbation-based method to  theory to derive a reduction method from infinite-dimensional to finite-dimensional systems. Building upon this work,
Ref. \cite{r4} has proposed a singular perturbation method to approximate a linear quantum system using its slow component whose physical realizability can be preserved.
To improve the accuracy of reduced models in the low-frequency regime, a truncation of a cascade cavity realization method is also developed for linear quantum systems \cite{r5}.
Subsequently, a quasi-balanceable truncation method has been introduced, which shows that all asymptotically stable completely passive linear
quantum systems have a quasi-balanced realization, and
an upper bound on the $\mathscr{H}_\infty$ error of the reduced model is obtained for the truncation \cite{r6}. Additionally, Ref. \cite{r7} has proposed a tangential interpolatory projection method for linear quantum systems to ensure accuracy at specific frequency points.
The above method is developed for quantum stochastic differential equations in the Heisenberg picture, another class of methods for model reduction is done in the Schr${\rm \ddot{o}}$dinger picture. Ref. \cite{chase2009} has obtained a reduced-dimensional description of quantum maps for collective quantum systems governed by Markovian master equations.
Ref. \cite{ticozzi2023} has introduced an operator projection method which constructs an effective subspace with respect to reachable sets of observables to find minimal linear realizations so as to reduce the dynamics of discrete-time quantum walks. This method can be extended to model reduction for continuous-time Markovian master equations~\cite{ticozzi2024b} , discrete-time conditional quantum systems under generalized measurements~\cite{ticozzi2024}, and continuous-time quantum filters~\cite{ticozzi2025}. Similarly, Ref. \cite{fan2024} has applied the principle of measurement-adapted coarse-graining to derive an effective quantum master equation.
%For general, Ref. has proposed an observable-projection-based model reduction method to derive minimal-dimensional approximations, which can be improved when a system has symmetric properties. Similarly, this method can be extended to.
%Regarding, Ref. has leveraged results from quantum probability and non-commutative conditional expectation theory to propose a method for computing minimal linear realizations of filters.
Collectively, although the above model reduction methods work for closed or Markovian quantum systems, the question of how to reduce the dimensions of non-Markovian quantum systems is still an open question.

  Non-Markovian quantum systems refer to a quantum system coupling to an environment with memory effects which lead to quite different dynamics from those of Markovian quantum systems~\cite{nm2,nm5,nm7,nm9,nm10}. Many approaches have been presented to describe non-Markovian dynamics of quantum systems, including time-convolution-less time-varying master equation~\cite{br}, integral-differential master equation~\cite{alireza} or Langevin equation~\cite{nm9,qip,weimin zhang}. In these equations, time-varying and integral terms are introduced to describe memory-effect-induced dynamics. However, these terms lead to challenges in designing filters or controllers for non-Markovian quantum systems. Hence, a class of augmented system models for non-Markovian quantum systems is proposed, where ancillary systems are introduced to represent the internal modes of a non-Markovian environment~\cite{nm10,nm11,nmCDC,nmQIP}.
  However, the introduction of the ancillary system would increase the dimensions of the augmented system.
  Hence, it is useful to consider model reduction methods to improve the computational efficiency of the augmented system model.

%  many practical quantum systems exhibit non-Markovian behavior, where the dynamics are intricately linked to the system's historical interactions with its environment \cite{nm1,nm2,nm3,nm4}. Such non-Markovian systems are often high-dimensional, posing significant computational challenges that hinder their analysis and control. Despite this critical need, model reduction for non-Markovian quantum systems remains an uncharted territory, highlighting the importance of developing efficient techniques tailored to these complex systems.

%Model reduction for non-Markovian quantum systems differs significantly from that for Markovian systems. Non-Markovianity typically arises from the memory effects of the environment, which complicate the system's dynamical behavior and make it difficult to describe using traditional Markovian models \cite{nm5,nm6,nm7,nm8}. To effectively model non-Markovian quantum systems, the augmented system approach is widely adopted, where ancillary system is introduced to capture the memory effects of the environment \cite{nm9,nm10,nm11}.
% However, this approach introduces new challenges: during the model reduction process, it is necessary to ensure not only the realizability and accuracy of the reduced system's dynamical properties but also the preservation of the principal system's specific structure.

%  Thus, model reduction for non-Markovian quantum systems requires balancing dimensionality reduction with the maintenance of complex dynamical properties and structural constraints, posing higher demands on model reduction techniques.

In this paper, we propose an $\mathscr{H}_2$ model reduction method for reducing the dimension of the augmented system for linear non-Markovian quantum systems. Concretely, we focus on reducing the dimension of the ancillary system since the principal system is of interest in general which should be preserved in the model reduction process.
Although current methods enable the design of ancillary systems to approximate environmental spectral functions, there remains a lack of comprehensive optimization  methods that balance the accuracy of the environmental representation with the dimensionality of the ancillary system from the perspective of the overall augmented system transfer function performance.
In addition, since the reduced model also describes an augmented system model of the non-Markovian quantum system, we give  physical realizability conditions for non-Markovian augmented system models.
 Different from traditional physical realizability for Markovian quantum systems, the ancillary-principal-system direct coupling structure is combined in the conditions.  Subject to these nonlinear conditions, an optimization problem for model reduction of the augmented system model is formulated, which is to minimize the $\mathscr{H}_2$ norm of the difference between the transfer functions of the original and reduced models. Necessary conditions are  derived for obtaining the input matrix and the system matrix in the reduced model.
 Through relaxation of nonlinear equality constraints to inequalities, the proposed optimization problem admits an efficient semidefinite programming solution.
 As a result, we obtain a reduced-order augmented system model for linear non-Markovian quantum systems while satisfying the corresponding physical realizability.

This paper is organized as follows. Section \ref{sec2} briefly reviews general augmented system models for non-Markovian quantum systems. Section \ref{sec3} describes the augmented system model for linear non-Markovian quantum systems and its corresponding physical realizability is proposed. Section \ref{sec4} formulates the optimization problem for model reduction of the augmented system model of linear non-Markovian quantum systems, where we also proposes our $\mathscr{H}_2$ model reduction method for the augmented system model.
Examples are given in Section \ref{sec5} to demonstrate the effectiveness of our method. Finally, conclusions are drawn in Section \ref{sec6}.

%
%
%Recognizing the limitations of existing methods like model truncation and projection, which preserve physical realizability but fall short of optimality, this paper introduces a novel model reduction strategy based on achieving the optimal H2 norm, offering a significant advancement in managing and controlling high-dimensional quantum   systems effectively.

\textbf{Notation} For a matrix $A=[A_{s,t}]$, the symbols $A^T$, $A^*$, and $A^\dag$ represent the transpose, conjugate, and conjugate transpose of $A$. $\mathcal{M}(A)_{s,t}=\frac{1}{2}\left[\begin{array}{cc}A_{s,t}+A_{s,t}^*&i(A_{s,t}-A^*_{s,t})  \\-i(A_{s,t}-A_{s,t}^*) &A_{s,t}+A_{s,t}^*\\\end{array}\right]$. Given two operators $M$ and $N$, their commutator $[\cdot,\cdot]$ is calculated as $[M,N]=MN-NM$.
%The matrix $\mathbb{J}_n$ is defined as $\mathbb{J}_n = I_n \otimes \begin{bmatrix}0 & 1 \\ -1 & 0\end{bmatrix}$.

\section{Brief Review of Augmented System Models of Non-Markovian Quantum Systems}\label{sec2}
\subsection{Description for Markovian Quantum Systems}
Generally, a Markovian quantum system refers to a quantum system interacting with quantum white noise and the variation of its state depends on its current state. Quantum white noise $b(t)$ is defined on a Fock space $\mathcal{F}$ satisfying the commutation relation $[b(t), b^\dagger(t')]=\delta(t-t')$, $[b(t), b(t')]=0$; i.e., there is no correlation between quantum white noises at two different instants~\cite{white}.

Supposing that a quantum system is defined on a Hilbert space $\mathcal{H}$, the evolution of the Markovian quantum system on the space $\mathcal{H}\otimes\mathcal{F}$ can be described by a quantum stochastic differential equation (QSDE)
\begin{equation}
\label{du}
dU(t)=\{-(iH+\frac{1}{2}L^\dagger L)dt+dB^\dagger_t L-L^\dagger dB_t\}U(t),
\end{equation}
where $U(t)$ is the evolution operator for the Markovian quantum system. The Hamiltonian of the system and the coupling operator between the system and the quantum white noise are denoted as $H$ and $L$, respectively. In addition,  the operator $B(t)=\int^t_{0}b(\tau)d\tau$ satisfying
$[B(t),B^\dag(\tau)]=\text{min}(t,\tau)$ and $\ [B(t),B(\tau)]=0$
 is a Wiener process. The quantum infinitesimal increments of $B(t)$ and $B^\dag(t)$ in a vacuum state satisfy the Ito rules $dB(t)dB^\dag(t)=dt$, $dB^\dag (t)dB(t)=0$, $dB(t)dB(t)=0$, and $dB^\dag(t)dB^\dag (t)=0$~\cite{r8}.

In the Heisenberg picture, the evolution of an operator $X$ of the system is expressed as $j(X)=U^\dagger(t) X U(t)$ which yields
\begin{equation}
dj(X)=j(\mathcal{G}(X))dt+dB_t^\dagger j([X, L])+j[L^\dagger, X]dB_t,\label{evol}
\end{equation}
where the generator $\mathcal{G}(X)$ is calculated as $
\mathcal{G}(X)=-i[X,H]+\frac{1}{2}L^\dag [X,L]+\frac{1}{2}[L^\dag, X]L.$

Note that in Eq.~(\ref{evol}), the white noise field is taken as the input of the system. After it interacts with and passes through the system, the field leaves the system, which is the output field. Mathematically, the output field is defined as $B_{out} (t) = U^\dag (t) B (t) U (t)$ so that the quantum infinitesimal increment for the output field satisfies
 \begin{equation}
dB_{out}(t)=j(L)dt+dB(t),\label{IO}
\end{equation}
which is the input-output relation of the Markovian quantum system~\cite{r8}.

Generally,  a Markovian quantum system,  can be described by the SLH description $(S, L, H)$, where the scattering matrix $S$ describes the scattering process of the fields. When the triple is given, the evolution and the input-output relation can be determined. For more details see Ref.~\cite{direct}.

\subsection{General Augmented System Model for Non-Markovian Quantum System}
Different from Markovian quantum systems, a non-Markovian quantum system refers to a quantum system interacting with quantum colored noise whose shaped spectrum is described by a non-Delta correlation function~\cite{nm9}. Inspired by shaping filters in classical control theory, a class of augmented system models for non-Markovian quantum systems has been presented in Ref.~\cite{nm10}. The main idea of the augmented system model is to take the states of spectrum-characterized internal modes of a non-Markovian environment into account such that a non-Markovian quantum system can be described in an augmented state space. The advantage of this model is the avoidance of time-varying and integral terms for the description of memory effects which benefits the design of filters and controllers.

To capture the internal dynamics of the non-Markovian environment described by a shaped spectrum, an ancillary system is introduced with Hamiltonian $H_a$ defined on a Hilbert space $\mathcal{H}_a$. To generate the required spectrum, the ancillary system is driven by quantum white noise defined on a Fock space $\mathcal{F}_a$, where their couplings are described by a coupling operator $L_a$ on the space $\mathcal{H}_a$. Before coming into the ancillary system, the quantum white noise involves no scattering process resulting in $S_a=I$. In terms of the SLH description, the ancillary system $\mathbf{G}_a$ can be described as
\begin{equation}
\mathbf{G}_a=(I, L_a, H_a).\label{ASa}
\end{equation}

Also, in the augmented system model, the quantum system of interest in the non-Markovian environment is named as the principal system, whose Hamiltonian is $H_p$ defined on a Hilbert space $\mathcal{H}_p$. It can interact with a field defined on a Fock space $\mathcal{F}_p$ for probing or connecting to other quantum systems. The couplings to the fields are denoted by an operator vector $L_p$ in the Hilbert space $\mathcal{H}_p$. Similarly, we consider that before interacting with the quantum system the fields involve no scattering process; i.e., $S_p=I$. Hence, the principal system $\mathbf{G}_p$ can be described by an SLH description as
\begin{equation}
\mathbf{G}_p=(I, L_p, H_p).\label{ASp}
\end{equation}

Since a non-Markovian quantum system exchanges energy with its environment in both directions, a direct interaction Hamiltonian $H_{pa}$ is introduced. In this way, the augmented system model SLH $\mathbf{G}$ is written as
\begin{equation}
\mathbf{G}=\left(
             \begin{array}{ccc}
              I,
                & \left(
                    \begin{array}{c}
                      L_p \\
                      L_a\\
                    \end{array}
                  \right),
                 & H_p+H_a+H_{pa} \\
             \end{array}
           \right)\label{GT}
\end{equation}
 and it evolves on a tensor product space $\mathcal{H}_p\otimes\mathcal{H}_a\otimes\mathcal{F}_p\otimes\mathcal{F}_a$. We have not specified the Hamiltonian or coupling operator of the system so that both the principal and ancillary systems can be atoms or linear quantum systems. Hence, Eq. (\ref{GT}) is a general SLH description for the augmented system  model of a non-Markovian quantum system. For more details, see Ref.~\cite{nm11}.

Note that the augmented system model is driven by quantum white noise so that its dynamics are Markovian. However, the principal system has non-Markovian dynamics due to the direct interaction with the ancillary system. Compared to existing models defined on $\mathcal{H}_p\otimes\mathcal{F}_p$, the dimension of the augmented system space may be very large in general, which would limit our ability to design a controller based on the augmented system model.

\section{Augmented System Models of Linear Non-Markovian Quantum Systems}\label{sec3}
\subsection{Hamiltonians and Coupling Operators for  Augmented System Model of Linear Non-Markovian Quantum Systems}
In this paper, we will focus on the augmented system models for linear non-Markovian quantum systems, where both the principal and ancillary systems are linear quantum systems.
The linear quantum systems for the ancillary system can generate an output with a rational spectrum, which can realize the dynamics of the internal modes of a non-Markovian environment.
Note that in the care that the noise spectrum of the non-Markovian environment is irrational, we can approximate it with a rational spectrum using Pad$\rm \grave{e }$ approximation. Hence, we can specify the principal and ancillary systems as follows.

For an $m$-mode linear principal system, the Hamiltonian is written as a quadratic form
\begin{equation}\label{hp}
 H_p=a_p^\dagger\Omega_pa_p,
\end{equation}
where the operator vector $a_p=\left[
                                                     \begin{array}{cccc}
                                                       a_{p1} & a_{p2} & \cdots & a_{pm}\\
                                                     \end{array}
                                                   \right]^T
$ is a vector of annihilation operators of the principal system. $a_p^\dagger=\left[
                                                                                    \begin{array}{cccc}
                                                                                      a_{p1}^\dagger & a_{p2}^\dagger & \cdots  & a_{pm}^\dagger  \\
                                                                                    \end{array}
                                                                                  \right]
$ is a row vector of creation operators of the principal system. The creation and annihilation operators satisfy the canonical commutation relations as
\begin{equation}\label{chcom}
  [a_{p},a_{p}^\dagger]=I_m, [a_{p},a_{p}]=0.
\end{equation}
In addition, the diagonal elements of the matrix $\Omega_p\in\mathbb{C}^{m\times m}$  determine the angular frequency of each mode and its off-diagonal elements describe the direct coupling strength of every pair of different modes. Also, we consider that the principal system is driven by $m'$ channels of probing fields so that the coupling operator can be written as
\begin{equation}\label{lp}
 L_p=N_pa_p,
\end{equation}
where $N_p\in\mathbb{C}^{m'\times m}$ is determined by the dissipation rate of the principal system to the probing fields.

Similarly, we consider that the ancillary system is an $n$-mode linear quantum system with the Hamiltonian
\begin{equation}\label{ha}
 H_a=a_a^\dagger\Omega_aa_a,
\end{equation}
where a vector of their annihilation operators is written as $a_a=\left[
                                                     \begin{array}{cccc}
                                                       a_{a1} & a_{a2} & \cdots & a_{an}\\
                                                     \end{array}
                                                   \right]^T
$ and correspondingly  a row vector of creation operators is denoted as $a_a^\dagger=\left[
                                                                                    \begin{array}{cccc}
                                                                                      a_{a1}^\dagger & a_{a2}^\dagger & \cdots  & a_{an}^\dagger  \\
                                                                                    \end{array}
                                                                                  \right]
$. Their components also satisfy the canonical commutation relations as
%\begin{equation}\label{chcom2}
%  [a_{aj},a_{ak}^\dagger]=\delta_{jk}, [a_{aj},a_{ak}]=0, j,k=1,\cdots,n.
%\end{equation}}
\begin{equation}\label{chcom2}
  [a_{a},a_{a}^\dagger]=I_n, [a_{a},a_{a}]=0.
\end{equation}
Also, the elements of $\Omega_a\in\mathbb{C}^{n\times n}$ are determined in an identical way as those for the principal system. In addition, since the ancillary system is used to capture the  spectrum of a non-Markovian environment, the ancillary system is driven by a quantum white noise process, where their couplings are characterized by the coupling operator
\begin{equation}\label{la}
 L_a=N_aa_a,
\end{equation}
with $N_a\in\mathbb{C}^{n\times n}$. Corresponding to this quantum white noise, a fictitious output is defined in the space $\mathcal{H}_a$ as
\begin{equation}\label{ficop}
  c=G_aa_a.
\end{equation}
with $G_a\in\mathbb{C}^{1\times n}$.
It can be checked that the fictitious output $c$ has a rational spectrum determined by the transfer function of the ancillary system from the input to the output field. Note that in this paper, we limit our discussion a single channel of fictitious outputs. More general cases will be discussed in future works.
%for the ancillary system with $G_a=[\sqrt{\gamma_1},\sqrt{\gamma_2},...,\sqrt{\gamma_n}]\in\mathbb{C}^{1\times n}$.

Moreover, the principal system is coupled to the ancillary system via a direct interaction Hamiltonian
\begin{equation}\label{hpa}
 H_ {pa} = i (c ^\dag z-z ^\dag c),
\end{equation}
where $z$ defined on Hilbert space $\mathcal{H}_p$, is a combination of operators of the principal system
\begin{equation}\label{diop}
  z=K_pa_p.
\end{equation}
with $K_p=[\sqrt{\kappa_1},\sqrt{\kappa_2},...,\sqrt{\kappa_m}]\in\mathbb{C}^{1\times m}$.
 In this way, the Hamiltonians and coupling operators are given such that we can obtain QSDEs of the augmented system model.

\subsection{QSDEs for the Augmented System Model of a Linear Non-Markovian Quantum System}
In the SLH description, the augmented system model can be described by \eqref{GT} with specified Hamiltonians \eqref{hp}, \eqref{ha} and \eqref{hpa} and coupling operators \eqref{lp} and \eqref{la} so that we can directly obtain the dynamics of the augmented system model which is given by the following Lemmas.

\begin{lemma} \label{Lemma1}\cite{nm10}: Consider the augmented model for the linear non-Markovian quantum system \eqref{GT} with specified Hamiltonians \eqref{hp}, \eqref{ha} and \eqref{hpa} and coupling operators \eqref{lp} and \eqref{la}. The system equation and the output equation of the augmented model of the linear non-Markovian quantum system can be expressed as
%Hence, with the SLH description for the augmented system, the quantum stochastic differential equation can be given  as
\begin{align}%\label{dx11}
{\rm d} a(t)&=Fa(t)dt+G{\rm d} B(t),\label{dx11}\\
{\rm d} B_{out}(t)&=Ha(t)dt+[I_{m\times m}~~0]{\rm d} B(t)\label{dx12}
\end{align}
with
\begin{align}
  F &=\left[ \begin{array}{cc}
            F_p & K_p^\dagger G_a \\
            -G_a^\dagger K_p & F_a \\
          \end{array}
        \right], \\
  G &= \left[  \begin{array}{cc}
                                   -N_p^\dagger & 0_{m\times n} \\
                                   0_{n\times m} & -N_a^\dagger \\
                                 \end{array}
                               \right], \\
  H &=\left[  \begin{array}{cc}
                                   N_p & 0_{m\times n}
                                 \end{array}
                               \right],
\end{align}
where $a(t)=\left[\begin{array}{c}
a_p(t) \\
a_a(t) \\
  \end{array}
\right]$ is the  state vector of annihilation operators. ${\rm d} B(t)=\left[
                                        \begin{array}{c}
                                          {\rm d} B_p(t)\\
                                          {\rm d} B_a(t)\\
                                        \end{array}
                                      \right]$ is the input field of the augmented system. ${\rm d} B_{out}(t)$ is the output field of the principal system, so the input-output matrix is given as $[I_{m\times m}~~0]$. The  submatrices $F_p$ and $F_a$ can be given as $F_p=-i\Omega_p-\frac{1}{2}N_p^\dagger N_p$,  $F_a=-i\Omega_a-\frac{1}{2}N_a^\dagger N_a$.
\end{lemma}
Notably, since both the principal system and the ancillary system constituting the augmented system are passive, the resulting augmented system retains passivity.
From Lemma \ref{Lemma1}, it is straightforward to obtain the following result.
\begin{lemma} \cite{direct}: \label{ff12}
For two directly coupled linear quantum passive systems described by the direct interaction Hamiltonian \eqref{hpa}, the off-diagonal submatrices of the system matrix $F$ in the quantum stochastic differential equation (QSDE) representation must satisfy the following relationship
\begin{align}
F_{12}=F_{12}^*,~F_{12}=-F_{21}^\dag.
\end{align}
\end{lemma}
However, since Eqs. (\ref{dx11}) and (\ref{dx12}) involve complex coefficients, it would be convenient to convert them into a quadrature representation with real coefficients.
In the quadrature representation, Eqs. (\ref{dx11}) and (\ref{dx12}) are rewritten as
\begin{align}
   {\rm d}x(t)&=Ax(t) {\rm d}t+B {\rm d}w,\nonumber\\
   {\rm d}y(t)&=Cx(t){\rm d}t+D{\rm d}w,\label{dx2}
\end{align}
where
\begin{align}A&=\left[ \begin{array}{cc}
            \bar F_{p} & \bar F_{pa}\\
            -\bar F_{pa}^T & \bar F_{a} \\
          \end{array}
        \right]=\left[ \begin{array}{cc}
            \mathcal{M}(F_p) & \mathcal{M}(K_p^\dagger G_a )\\
            \mathcal{M}(-G_a^\dagger K_p) & \mathcal{M}(F_a )\\
          \end{array}
        \right],\nonumber\\
        B&=\left[ \begin{array}{cc}
            \bar G_{p} & 0_{2m\times 2n}\\
            0_{2n\times 2m} & \bar G_{a} \\
          \end{array}
        \right]=\left[  \begin{array}{cc}
                                   \mathcal{M}(-N_p^\dagger) & 0_{2m\times 2n} \\
                                   0_{2n\times 2m} & \mathcal{M}( -N_a^\dagger ) \\
                                 \end{array}
                               \right],\nonumber\\
C&=\left[ \begin{array}{cc}
            -\bar G_{p}^T & 0_{2m\times 2n}
          \end{array}
        \right],\  D=\left[ \begin{array}{cc}
            I_{2m\times 2m} & 0
          \end{array}
        \right], \label{ABC}\end{align}
with the operation $\mathcal{M}(\cdot)$ defined in the notation of Section \ref{sec1}.
The submatrices $\bar F_p\in \mathbb{R}^{2m\times2m}$ and $\bar F_a\in\mathbb{R}^{2n\times2n}$ correspond to the principal and the ancillary system, respectively. The submatrix $\bar F_{pa}\in\mathbb{R}^{2m\times2n}$ characterizes the couplings between the principal and the ancillary systems.
The state vector is denoted as $x(t) =[
               q_{p_1}(t), p_{p_1}(t),..., q_{p_m}(t), p_{p_m}(t), q_{a_1}(t) , p_{a_1}(t),..., q_{a_n}(t),$ $ p_{a_n}(t)]^T$, where $q_{p_j}(t)$ and $p_{p_j}(t)$ are the position and momentum operators of the principal system, respectively;
$q_{a_j}(t)$ and $p_{a_j}(t)$ are the position and momentum operators of the ancillary system, respectively.
The input vector can be rewritten as           ${\rm d}w(t)= [
             {\rm d}v_{pq_1}(t), {\rm d}v_{pp_1}(t),..., {\rm d}v_{pq_m}(t), {\rm d}v_{pp_m}(t),{\rm d}v_{aq_1}(t), {\rm d}v_{ap_1}(t)$ $,...,{\rm d}v_{aq_n}(t), {\rm d}v_{ap_n}(t)
            ]^T$ with $[
             {\rm d}v_{pq_j}(t)~~{\rm d}v_{pp_j}(t)]=\frac{1}{\sqrt 2}\left[ \begin{array}{cc}
            1 & 1\\
            -i & i \\
          \end{array}
        \right] [{\rm d} B_{pj}(t)~~{\rm d} B_{pj}^\dag(t)]^T$ and $[
             {\rm d}v_{aq_j}(t)~~ {\rm d}v_{ap_j}(t)]=\frac{1}{\sqrt 2}\left[ \begin{array}{cc}
            1 & 1\\
            -i & i \\
          \end{array}
        \right][{\rm d} B_{aj}(t)~~{\rm d} B_{aj}^\dag(t)]^T$, where ${\rm d} B_{pj}(t)$ and ${\rm d} B_{aj}(t)$ are the $j_{th}$ input field for principal and ancillary systems, respectively.
              The transfer function of the system is
            $$\Xi_{G}(s)=C(sI_{(2m+2n)\times(2m+2n)}-A)^{-1}B+[I_{2m\times2m}~~0].$$
            We will evaluate the performance of model reduction via the difference between the transfer functions of the original and reduced system models.
\subsection{Physical Realizability of the Augmented System Model}
Generally, the coefficient matrices of QSDEs should satisfy specific conditions which guarantee that the QSDEs describe a genuine quantum system. These conditions are referred to physical realizability conditions~\cite{ccr}. Here, we give some results on physical realizability for the augmented system model for linear non-Markovian quantum systems.

\textbf{Definition 1.} (Physical realizability for the augmented system model)
A system described by linear QSDEs of the form in Eqs.~(\ref{dx11}) and (\ref{dx12})
is said to be a
physically realizable augmented system model for linear non-Markovian quantum systems if the following conditions hold:

\begin{itemize}
    \item[(i)] The  system  described by Eqs. (\ref{dx11}) and (\ref{dx12}) is Markovian and physically realizable;
    \item[(ii)]The state vector can be divided into two parts corresponding to two subsystems and the corresponding matrices have block structures with suitable dimensions. The two subsystems are directly coupled.
    \item[(iii)] The fields interacting with the system in Eq. (\ref{dx11}) can also be divided into two parts interacting with the two subsystems, respectively.
\end{itemize}

%given augmented system composed of two subsystems, one of the subsystems is said to be non-Markovian if
%\begin{itemize}
%    \item[(i)] The passive augmented Markovian system  is physically realizable;
%    \item[(ii)] The two subsystems are directly coupled.
%\end{itemize}

Given Definition 1, it is necessary to give conditions on the matrices in Eqs.~(\ref{dx11}) and (\ref{dx12}) to determine the physical realizability of the augmented system model for linear non-Markovian quantum systems. Then, we can give the following theorem.

\begin{theorem} \label{th1} The augmented system model for linear non-Markovian quantum systems described by Eqs. \eqref{dx11} and (\ref{dx12})
 with $F=\left[ \begin{array}{cc}
           F_{11} & F_{12}\\
            F_{21} & F_{22} \\
          \end{array}
        \right]$, $G=\left[ \begin{array}{cc}
           G_{11} & G_{12}\\
            G_{21} & G_{22} \\
          \end{array}
        \right]$, $H=\left[ \begin{array}{cc}
           H_{11} & H_{12}
          \end{array}
        \right]$,
 is physical realizable  if and only if
\begin{align}
F_{11} + F_{11}^\dag + G_{11}G_{11}^\dag &= 0, \label{PR1} \\
G_{11} &= -H_{11}^\dag, \label{PR2}\\
F_{22} + F_{22}^\dag + G_{22}G_{22}^\dag &= 0, \label{PR3} \\
%G_{22} &= -H_{22}^\dag, \label{PR4}\\
F_{12}&=F_{12}^*,\label{Re4}\\
F_{12}&=-F_{21}^\dag,\label{f12}\\
G_{12}&=H_{12}=0,\label{g12}
%G_{21}&=H_{21}=0. \label{g21}
\end{align}
\end{theorem}

\textbf{Proof.}~Sufficiency: it is straightforward to find that a system described by Eqs.~(\ref{dx11}) and (\ref{dx12}) is Markovian since the variation of the state depends on the current state and the inputs are quantum white noise.  Supposing \eqref{PR1}-\eqref{g12} hold, we have
 \begin{eqnarray}
&&F+F^\dag+GG^\dag\nonumber\\
&=&\left[ \begin{array}{cc}
           F_{11} & F_{12}\\
            F_{21} & F_{22} \\
          \end{array}
        \right]+\left[ \begin{array}{cc}
           F_{11} & F_{12}\\
            F_{21} & F_{22} \\
          \end{array}
        \right]^\dag\nonumber\\
        &&~~~~~~~~~~~~~~~~~+\left[ \begin{array}{cc}
           G_{11} & G_{12}\\
            G_{21} & G_{22} \\
          \end{array}
        \right]\left[ \begin{array}{cc}
           G_{11} & G_{12}\\
            G_{21} & G_{22} \\
          \end{array}
        \right]^\dagger,\nonumber\\
        &=&\left[ \begin{array}{cc}
           F_{11}+ F_{11}^\dagger+ G_{11} G_{11}^\dag& F_{12}+F_{21}^\dag\\
            F_{21}+F_{12}^\dag & F_{22}+ F_{22}^\dagger + G_{22} G_{22}^\dag\\
          \end{array}
        \right]\nonumber\\
        &=&0 \label{fg}.
%G&=&-H^\dag,\label{bc}
\end{eqnarray}
Similarly, with \eqref{PR2} and \eqref{g12}, we have
\begin{equation}\label{gh}
  G\left[ \begin{array}{c}
            I \\ 0
          \end{array}
        \right]=\left[ \begin{array}{cc}
           G_{11} & G_{12}\\
            G_{21} & G_{22} \\
          \end{array}
        \right]\left[ \begin{array}{c}
            I \\ 0
          \end{array}
        \right]=\left[ \begin{array}{c}
           -H_{11}^\dag \\
           -H_{12}^\dag \\
          \end{array}
        \right]=-H^\dag,
\end{equation}
According to Ref. \cite{ccr}, since \eqref{fg} and \eqref{gh} hold, the system described by Eqs.~(\ref{dx11}) and (\ref{dx12})  is physically realizable.
From Lemma \ref{ff12} and condition \eqref{f12}, the two subsystems of the augmented system model are directly coupled.
Then, due to \eqref{PR2} and \eqref{g12}, we find the fields interact with the two subsystems, respectively, which involve no scattering.
Therefore, \eqref{PR1}-\eqref{g12}  are sufficient for physical realizability.

Necessity: Conversely, if the system described by Eqs. \eqref{dx11} and (\ref{dx12}) is physically realizable and the state can be divided into two parts corresponding to two subsystems, respectively, the two subsystems are also physically realizable so that \eqref{PR1}-\eqref{PR3} hold. From Definition 1 (ii) and (iii), we get \eqref{Re4}-\eqref{g12} directly.
$\blacksquare$

Note that the augmented system model \eqref{dx11} is Markovian, which satisfies the physical realizability conditions for Markovian quantum systems~\cite{ccr}. However, due to the direct interactions between the two subsystems, the conditions in Theorem \ref{th1} are stronger than those for general Markovian quantum systems.

For convenience, we also give the following physical realizability conditions for  the system \eqref{dx2} in the quadrature representation.

\begin{corollary}\label{coro}  The augmented system model for linear non-Markovian quantum systems described by Eq. \eqref{dx2} is physically realizable  if and only if the following conditions are satisfied:
         \begin{align}
A_{11}+A_{11}^T+B_{11}B_{11}^T&=0,\label{real21}
\\
 B_{11}&=-C_{11}^T,\label{real22}
\\
    A_{22}+A_{22}^T+B_{22}B_{22}^T&= 0 ,\label{real23}
%\\
%    B_{22}&=-C_{22}^T,\label{real24}
    \\
         A_{12}& = K_G \otimes I_{2\times2},\label{KG}\\
         A_{12}&=-A_{21}^T, \label{real25}
    \\
    B_{12}=C_{12}&=0,
%    \\
%    B_{21}=C_{21}&=0,
     \label{real26}
\end{align}
%         \begin{eqnarray}
%\bar F_p+\bar F_p^T+\bar G_p\bar G_p^T&=&0,\label{real21}
%\\
%\bar F_a+\bar F_a^T+\bar G_a\bar G_a^T&=& 0,\label{real22}
%\\
%     \bar F_{pa} - K_G \otimes I_{2\times2}&=&0, \label{real23}
%\\
%     \textcolor{red}{B+C^T}&=&0. \label{real25}
%\end{eqnarray}
where $K_G\in \mathbb{R}^{m\times n}$ is related to the coupling strength.
\end{corollary}

\section{Model Reduction for the  Augmented System Model of Linear Non-Markovian Quantum Systems }\label{sec4}

In the above augmented system model, to accurately capture the non-Markovian dynamics, it is necessary to introduce a sufficient number of ancillary systems which would increase the dimension of the augmented system, drastically resulting in a heavy computational burden when using the model for controller design.
In this paper, to reduce the dimension of the augmented system, we explore a model reduction method to obtain a system model with low dimensions to approximate the original system. As we have introduced, we expect to realize it under a criterion which can guarantee the performance of the obtained model.

\subsection{Formulation of Model Reduction of the Augmented System Model as an Optimization Problem}

\subsubsection{Description of the Reduced Model}

\begin{figure}[htbp]
\centerline{\includegraphics[width=8cm]{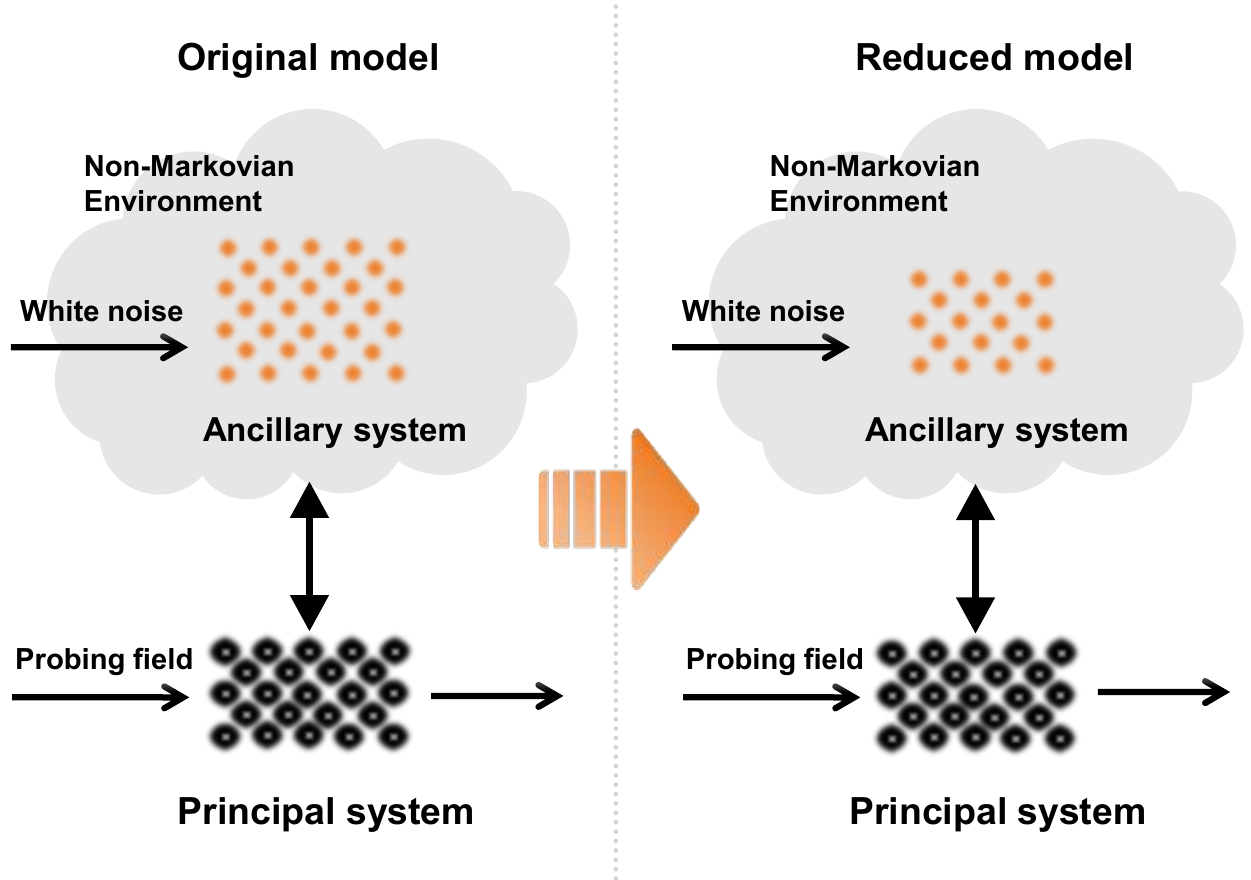}}
\caption{Schematic Diagram of Model Reduction for Augmented Systems for Linear Non-Markovian Quantum Systems }
\label{fig0}
\end{figure}

% When we consider the model reduction problem in non-Markovian linear quantum system models, we only consider reducing the number of modes of the ancillary system in model \eqref{dx2}, while the dynamics of principal system is preserved. Since the principal system is only coupling with the first oscillator of the ancillary system,
% the number of channels of input and output fields are kept the same as those of the original system \eqref{dx2}.
%So we write the state vector of a reduced system as $x_r(t)\in\mathbb R^{2+2r}$ with  $r<n$, which is in a similar form of $x(t)$  and the dynamical equation of the reduced model is described as
In model reduction, we focus on reducing the number of modes of the ancillary system in Eq. \eqref{dx2}, while we keep the principal system unchanged, since the principal system is of interest in general, as shown in Fig. \ref{fig0}. In addition, since the input fields of the augmented system have their functions, we consider the number of input fields remains the same as those of the original system \eqref{dx2}. Similarly, the number of output fields in the reduced model is preserved to match that of the original system, ensuring consistency in measurement accessibility. For example, as we mentioned, the input field of the ancillary system drives to generate quantum colored noise and the input fields of the principal system would be used for probing. However, the output fields $y_r(t)$ in the reduced model would be different from those in the original model due to the reduction of the state modes.

Hence, in the quadrature representation, the state vector of the reduced system is expressed as $x_r(t) =[
               q_{p_1}(t), p_{p_1}(t),..., q_{p_m}(t),  p_{p_m}(t), \bar q_{a_1}(t) ,\bar p_{a_1}(t),...,\bar q_{a_r}(t),$ $ \bar p_{a_r}(t)]^T$, with \( r < n \), which maintains a structure analogous to \( x(t) \). Since the input and output fields in the reduced model are identical to those in the original model, their symbols are also be kept. The QSDE of the reduced augmented system model is formulated as
\begin{align}\label{dxr}
&{\rm d}x_r(t)=A_rx_r(t){\rm d}t+B_r{\rm d}w(t),\nonumber\\
&{\rm d}y_r(t)=C_rx_r(t){\rm d}t+[I_{2m\times2m}~~0]{\rm d}w(t).
 \end{align}
Here, the matrices $A_r\in \mathbb{R}^{(2m+2r)\times(2m+2r)}$,  $B_r\in \mathbb{R}^{(2m+2r)\times(2m+2r)}$ and
$C_r\in \mathbb{R}^{2m\times(2m+2r)}$
are to be determined.
With the reduced model, it is straightforward to obtain the transfer function of the reduced system as
 $$\Xi_{G_r}(s)=C_r(sI_{2(m+r)\times2(m+r)}-A_r)^{-1}B_r+[I_{2m\times2m}~~~0].$$

\subsubsection{Construction of the Objective Function}

Different from existing methods typically using a metric to assess reduced models after model reduction, we adopt the $\mathscr{H}_2$ norm of the difference $\hat\Xi(s)$ between the original transfer function $\Xi_G(s)$ and the reduced model transfer function $\Xi_{G_r}(s)$; i.e.,
 \begin{align}\label{H2}
||\Xi_G(s)-\Xi_{G_r}(s)||_2 = \left( \frac{1}{2\pi} \int_{-\infty}^{\infty} \text{tr}\left[ \hat \Xi(j\omega)\hat \Xi^*(j\omega) \right] {\rm d}\omega \right)^{\frac{1}{2}}
\end{align}
to design the reduced model. We thus expect to solve the model reduction problem by optimizing the objective \eqref{H2} such that the performance of model reduction can be guaranteed.
However, the calculation of the  $\mathscr{H}_2$ norm involves the integration of the difference of the transfer functions over all frequencies, which is not convenient.
To calculate it in an alternative way, we augment the original and reduced systems as
\begin{align}
&{\rm d} \hat x(t)=\hat   A  \hat x(t){\rm d}t+  \hat B{\rm d}w(t),\nonumber\\
&{\rm d} \hat y(t)=  \hat C  \hat x(t){\rm d}t\label{dx3}
\end{align}
with
$
  \hat x(t)=\left[\begin{array}{c}x(t)  \\x_r(t)\\\end{array}\right]$, $
  \hat y(t)=y(t)-y_r(t)$,
and
  \begin{align}
  \hat A&=\left[\begin{array}{cc}A&0_{2(m+n)\times 2(m+r)}  \\0_{2(m+r)\times 2(m+n)} &A_r\\\end{array}\right],\nonumber\\
  \hat B&=\left[\begin{array}{c}B\\B_r\\
\end{array}\right],\nonumber\\
  \hat C&=\left[\begin{array}{cc}C &-C_r\\\end{array}\right]
.\nonumber
\end{align}

Hence, the following classical result can be adopted.

%\textcolor{blue}{should we mention controllability and observability
%\begin{lemma} \cite{hur}:
% For a passive linear system, Hurwitz stability, controllability and observability are equivalent.
%\end{lemma}}
%
%
%\textcolor{red}{The stability of the system \eqref{dx3} can be guaranteed since both the original system model \eqref{dx2} and the reduced model \eqref{dxr} are physical realizable. }
%\textcolor{blue}{I am not sure the above sentence expresses a correct logic.}
%Hence, the following classical result can be adopted.
%
%\textcolor{blue}{Since both the original system model \eqref{dx2} and the reduced model \eqref{dxr} are Hurwitz stable, the system \eqref{dx3} is also Hurwitz stable. Hence, we have the following Lemma.}

\begin{lemma} \label{lemma3} \cite{robust}
 If the total system \eqref{dx3} is Hurwitz stable, then, there exist positive definite matrices $P\in \mathbb{R}^{(4m+2n+2r)\times(4m+2n+2r)}$ or $  Q\in \mathbb{R}^{(4m+2n+2r)\times(4m+2n+2r)}$, which satisfy Lyapunov equations
\begin{align}
      \hat A   P +   P   \hat A^T +  \hat  B  \hat  B^T   = 0, \label{barP}
    \end{align}
    for $\hat B\hat B^T>0$, and
    \begin{align}
     \hat  A^T   Q +   Q   \hat A +  \hat  C^T   \hat C  = 0, \label{barQ}
\end{align}
for $\hat C^T\hat C>0$, respectively.
\end{lemma}
With the augmented system \eqref{dx3} and Lemma \ref{lemma3}, an alternative way to calculate the $\mathscr{H}_2$ norm is given in the following lemma.

\begin{lemma} \cite{H2} \label{lemma4}
The  square of the $\mathscr{H}_2$ norm \eqref{H2} can be calculated as
   \begin{align}
||\Xi_G(s)-\Xi_{G_r}(s)||^2_2={\rm tr}(  \hat C  P \hat  C^T).\label{XiG1}
\end{align}
or
   \begin{align}
||\Xi_G(s)-\Xi_{G_r}(s)||^2_2={\rm tr}(  \hat B^T  Q  \hat B).\label{XiG}
\end{align}
\end{lemma}
From Lemma \ref{lemma4}, the objective function of the optimization problem can be formulated.

\subsubsection{Constraints Formulation for Optimization}
To optimize the $\mathscr{H}_2$ norm for model reduction, we also have constraints. On the one hand, to obtain a Hurwitz-stable reduced model, we should first ensure a Hurwitz-stable augmented system. From Lemma~\ref{lemma3}, if we can find $P$ or $Q$ satisfying the Lyapunov equations, the augmented system is Hurwitz stable. Hence, in optimizing the $\mathscr{H}_2$ norm for model reduction, we should include the Lyapunov equations as constraints. On the other hand, the reduced model also represents an augmented system model for the original linear non-Markovian quantum systems, so  the corresponding matrices $A_r$, $B_r$, $C_r$ should satisfy the physical realizability conditions.
According to the physical realizability conditions in Corollary~\ref{coro}, the system matrices must have the following form:
 \begin{align}
A_r&=\left[ \begin{array}{cc}
            \tilde  F_{p} & \tilde  F_{pa}\\
            -\tilde  F_{pa}^\dag & \tilde  F_{a} \\
          \end{array}
        \right],\label{Ar1}\\
B_r&=\left[ \begin{array}{cc}
            \tilde  G_{p} & 0_{2m\times2r}\\
            0_{2r\times2m} & \tilde  G_{a} \\
          \end{array}
        \right],\label{Br1}\\
C_r&=\left[ \begin{array}{cc}
            -\tilde  G_{p}^T & 0_{2m\times2r}
          \end{array}\label{Cr1}
        \right],\end{align}
        and satisfy
\begin{align}
\tilde F_p+\tilde F_p^T+\tilde G_p\tilde G_p^T&=0,\label{realr21}\\
\tilde F_a+\tilde F_a^T+\tilde G_a\tilde G_a^T&=0,\label{realr22}\\
  \tilde F_{pa} - \tilde F_{m\times n} \otimes I_{2\times2}&=0.\label{realr23}
\end{align}
Here, the condition \eqref{real25} is obviously satisfied and thus omitted.

 Different from classical model reduction, the reduced model \eqref{dxr} not only minimizes the $\mathscr{H}_2$ norm  but also satisfies the physical realizability conditions which induce nonlinear constraints.

Since the order of the ancillary system is reduced, we have reduced-dimension submatrices $\tilde F_a\in\mathbb{R}^{2r\times2r}$, $\tilde F_p\in\mathbb{R}^{2m\times2m}$, $\tilde F_{pa}\in\mathbb{R}^{2m\times2r}$,
$\tilde G_a\in\mathbb{R}^{2r\times2r}$ and $\tilde G_p\in\mathbb{R}^{2m\times2m}$,  and the submatrix for the principal system is kept; i.e.,
% \textcolor{red}{$\tilde F_{p}=\bar F_p$, $\tilde G_{p}=\bar G_p$}. On the other hand, we should keep  the original principal system in the reduced model so that we have
  \begin{align}
\tilde F_p&=\bar F_p ,\label{pr1}\\
\tilde G_p&=\bar G_p.\label{pr3}
\end{align}

According to Eq. \eqref{pr3}, we find that  $C_r$ is no longer a matrix to be determined.

\subsubsection{Optimization for $\mathscr H_2$ model reduction of the augmented system model}

The model reduction problem for the augmented system model of a linear non-Markovian quantum systems can be summarized as:

%Given the augmented system of the linear non-Markovian quantum system~\eqref{dx2} satisfying the realizability conditions \eqref{real21}-\eqref{real25}, we design a reduced model in the form of Eq. \eqref{dxr}  satisfying the physically realizability conditions (\ref{realr21})-(\ref{realr23}), and conditions (\ref{pr1}) and (\ref{pr3}) such that  the $\mathscr{H}_2$ norm of the difference of the transfer functions is minimized; i.e.,
%\begin{equation}
%\begin{aligned}
%& \min_{A_r,B_r,C_r} J=\left\| \Xi_G(s) - \Xi_{G_r}(s) \right\|_2 \\
%& \text{s.t.} \quad \text{Eqs. (\ref{realr23}),~(\ref{pr1}) }.
%\end{aligned}
%\end{equation}

For the augmented system model of a linear non-Markovian quantum  system  \eqref{dx2} satisfying physical realizability conditions \eqref{real21}-\eqref{real26},  find a Hurwitz matrix $A_r$, a real-valued matrix $B_r$ and a positive definite matrix $P$ or $Q$ that minimizes the square of the $\mathscr{H}_2$ norm \eqref{XiG}; i.e,
\begin{align}
    & \min_{A_r, B_r,P, ~Q} J={\rm tr}(  \hat B^T  Q  \hat B)={\rm tr}(  \hat C  P \hat  C^T) \label{optp} \\
     &~~~~~~\text{s.t.}\ \rm{Eqs.}
     \eqref{barP}-\eqref{barQ},\ \eqref{Ar1}-\eqref{pr3}.\nonumber
\end{align}

\subsection{Two Necessary Conditions for Optimization of Model Reduction of the Augmented System Model}

%Therefore, we can transform the $\mathscr{H}_2$ model reduction problem into solving the following optimization problem.

We now derive the necessary conditions for solving the optimization problem \eqref{optp}.

\begin{theorem}\label{th4}
 For the original augmented system model \eqref{dx2}, if the model reduction problem \eqref{optp} is solvable and the positive definite matrix solution $Q$ of the Eq. \eqref{barQ} has the following block structure form
$Q=\left[\begin{array}{cc}  Q_1&  Q_2  \\  Q_2^T &  Q_3\\\end{array}\right]$, where $Q_2=\left[\begin{array}{cc}  Q_{211}&  Q_{212}  \\  Q_{221} &  Q_{222}\\\end{array}\right]$,
    $Q_3=\left[\begin{array}{cc}  Q_{311}&  Q_{312}  \\  Q_{312}^T &  Q_{322}\\\end{array}\right]$,
then the input matrix \( B_r \) in (\ref{Br1}) of the reduced augmented system model \eqref{dxr} is in the form of
  \begin{align}
  B_r&=\left[ \begin{array}{cc}
            \tilde  G_{p} & 0_{2m\times2r}\\
            0_{2r\times2m} & Q_{322}^{-1}Q_{222}^T\bar G_a \\
          \end{array}
        \right], \label{Br}
\end{align}
where $\bar G_a$ has been defined in \eqref{ABC}
and the following conditions should be satisfied
   \begin{align}
 Q_{211}^T+Q_{311}&=0, \label{211}\\
  Q_{212}+Q_{312}&=0,\label{212}\\
  Q_{221}^T+Q_{312}Q_{322}^{-1}Q_{222}^T&=0.\label{221}
\end{align}
\end{theorem}

 \textbf{Proof.}
Supposing that $\nu$ is an arbitrary element in $A_r$ or $B_r$ , the derivative of the cost function $J= \text{tr}( \hat CP\hat  C^T)$ with respect to $\nu$ is calculated as
   \begin{align}
\frac{\partial J}{\partial \nu}=\mathrm{tr}\left(\frac{\partial P}{\partial \nu}  \hat C^T  \hat C\right).\label{25-0}
\end{align}
Due to the difficulty in calculation of $\frac{\partial P}{\partial \nu}$, we first utilize Lyapunov equation \eqref{barQ} to rewrite Eq. \eqref{25-0} as
   \begin{align}\label{21J}
\frac{\partial J}{\partial \nu}=-2\mathrm{tr}\left(\frac{\partial P}{\partial \nu}Q  \hat A\right).
\end{align}
We then take the derivative of \eqref{barP} with respect to $\nu$ and obtain
   \begin{align}\label{25b}
\frac{\partial  \hat  A}{\partial \nu}P+  \hat A\frac{\partial P}{\partial \nu}+\frac{\partial P}{\partial \nu}  \hat A^T+P\frac{\partial  \hat  A^T}{\partial \nu}+\frac{\partial   \hat B \hat  B^T}{\partial \nu}=0.
\end{align}
Post-multiplying  \eqref{25b} by $Q$ and taking the trace, we obtain
   \begin{align} \label{23C}
2\mathrm{tr}\left(\frac{\partial \hat   A}{\partial \nu}PQ\right)+2\mathrm{tr}\left(\frac{\partial P}{\partial \nu}Q  \hat A\right)+\mathrm{tr}\left(\frac{\partial \hat   B  \hat B^T}{\partial \nu}Q\right)=0.
\end{align}
Hence, with Eqs. \eqref{21J} and \eqref{23C}, Eq. \eqref{25-0} can be rewritten as
%We finally obtain the following equation by combining \eqref{21J} and \eqref{23C}
   \begin{align}
\frac{\partial J}{\partial \nu}=
2\mathrm{tr}\left(\frac{\partial \hat   A}{\partial \nu}PQ\right)+\mathrm{tr}\left(\frac{\partial   \hat B \hat  B^T}{\partial \nu}Q\right).\label{Jv}
\end{align}
Hence, when $\nu$ is an arbitrary element of the matrix $B_r$, we  obtain
   \begin{align}
&\frac{\partial J}{\partial \nu}=
\mathrm{tr}\left(\frac{\partial  \hat  B  \hat B^T}{\partial \nu}Q\right)\nonumber\\
&=\mathrm{tr}\left(\left[\begin{array}{cc}0&B \frac{\partial B_r^T}{\partial \nu}\\\frac{\partial B_r}{\partial \nu}B^T &\frac{\partial B_r}{\partial \nu}B_r^T+B_r\frac{\partial B_r^T}{\partial \nu}
\\\end{array}\right]\left[\begin{array}{cc}Q_1&Q_2  \\Q_2^T &Q_3\\\end{array}\right]\right)\nonumber\\
&=\mathrm{tr}\left(B\frac{\partial B^T_r}{\partial \nu}Q_2^T+\frac{\partial B_r}{\partial \nu}B^TQ_2+\frac{\partial B_r}{\partial \nu}B^T_rQ_3+B_r\frac{\partial B_r^T}{\partial \nu}Q_3\right)\nonumber\\
&=2\mathrm{tr}\left(\frac{\partial B^T_r}{\partial \nu}(Q_2^TB+Q_3B_r)\right).\label{jnu}
\end{align}
When $J$ is minimized,  we have $Q_2^TB+Q_3B_r=0$.
This yields
   \begin{align}
Q_{211}^T\bar G_p+Q_{311}\bar G_p&=0,\label{Q211}\\
Q_{221}^T\bar G_a+Q_{312}\tilde G_{a}&=0,\\
Q_{212}^T\bar G_p+Q_{312}^T\bar G_p&=0,\\
Q_{222}^T\bar G_a+Q_{322}\tilde G_a&=0.\label{Q222}
\end{align}
Combining conditions \eqref{Q211}-\eqref{Q222}, we obtain  conditions \eqref{Br}-\eqref{211}.
$\blacksquare$

\textbf{Remark 1.}
Theorem \ref{th4} provides necessary conditions for the design of $B_r$, from which the matrix $C_r$ can be obtained.
From the structure of Eq. \eqref{Br}, the matrix corresponding to the principal system has been preserved.
 Based on this, our subsequent task focuses on the design of the matrix $A_r$ in the reduced model.

%Firstly, we define a $2r$-dimensional reduced model
%\begin{align}\label{dxs}
%&dx_s(t)=A_sx_s(t)dt+B_sdw(t),\nonumber\\
%&dy_s(t)=C_sx_s(t)dt+Ddw(t),
% \end{align}
% where $A_s\in \mathbb R^{2r\times2r}$, $B_s\in \mathbb R^{2r\times2m}$ and $C_s\in \mathbb R^{2m\times2r}$.
% The transfer function is $\Xi_{G_s}(s)=C_s(sI-A_s)^{-1}B_s+D$.
%Next, according to the system model \eqref{dx2} and the reduced model \eqref{dxs}, an augmented system can be defined as
%\begin{align}
%&    A=\left[\begin{array}{cc}A&0  \\0 &A_s\\\end{array}\right],\nonumber\\
%&    B=\left[\begin{array}{c}B\\B_s\\
%\end{array}\right],\nonumber\\
%&    C=\left[\begin{array}{cc}C &-C_s\\\end{array}\right].
%\end{align}
% Then we can
% convert optimization problem \eqref{opt} into the following form.

\begin{theorem}\label{th3} For the original augmented system model of a linear non-Markovian quantum system \eqref{dx2}, if the model reduction problem \eqref{optp} is solvable and the positive definite matrices $P$ and $Q$ have the following block structure form
$P=\left[\begin{array}{cc}  P_1&  P_2  \\  P_2^T &  P_3\\\end{array}\right]$,$P_2=\left[\begin{array}{cc}  P_{211}&  P_{212}  \\  P_{221} &  P_{222}\\\end{array}\right]$,
    $P_3=\left[\begin{array}{cc}  P_{311}&  P_{312}  \\  P_{312}^T &  P_{322}\\\end{array}\right]$, $Q=\left[\begin{array}{cc}  Q_1&  Q_2  \\  Q_2^T &  Q_3\\\end{array}\right]$, $Q_2=\left[\begin{array}{cc}  Q_{211}&  Q_{212}  \\  Q_{221} &  Q_{222}\\\end{array}\right]$,
    $Q_3=\left[\begin{array}{cc}  Q_{311}&  Q_{312}  \\  Q_{312}^T &  Q_{322}\\\end{array}\right]$,
then the system matrix \( \tilde F_a \) in (\ref{Ar1}) of the reduced augmented system model \eqref{dxr} is in the form of
  \begin{align}
\tilde F_a=-Q_{322}^{-1}Q_{222}^T \bar F_a P_{222}P^{-1}_{322}, \label{Ar}
\end{align}
where $\bar F_a$ is defined as in \eqref{ABC}
and the following conditions are satisfied:
   \begin{align}
 P_{212}&=0, \label{P212}\\
  P_{312}&=0.\label{P312}
\end{align}
\end{theorem}

 \textbf{Proof.} Using the result in Theorem \ref{th4},
for any element \(\nu\) in the matrix \(A_r\) of Eq. \eqref{Jv}, the partial derivative of \(J\) with respect to \(\nu\) satisfies
  \begin{align}
0&=\frac{\partial J}{\partial \nu}=
2\mathrm{tr}\left(\frac{\partial \hat A}{\partial \nu}PQ\right)\nonumber\\
&=2\mathrm{tr}\left(\left[\begin{array}{cc}0&0  \\0 &\frac{\partial A_r}{\partial \nu}\\\end{array}\right]\left[\begin{array}{cc}P_1&P_2  \\P_2^T &P_3\\\end{array}\right]\left[\begin{array}{cc}Q_1&Q_2  \\Q_2^T &Q_3\\\end{array}\right]\right)\nonumber\\
&=2\mathrm{tr}\left(\frac{\partial A_r}{\partial \nu}(P_2^TQ_2+P_3Q_3)\right).
\end{align}

Since \(\nu\) is an arbitrary entry, it follows that
\begin{align}
P_2^T Q_2 + P_3 Q_3 = 0. \label{P2Q2}
\end{align}
To determine \( \tilde F_a \), we expand Eq. \eqref{barP} and derive the following system of equations:
   \begin{align}
AP_2+P_2A_r^T+BB^T_r&=0,\label{AP2}\\
A_rP_3+P_3A^T_r+B_rB^T_r&=0.\label{ArP3}
\end{align}

It can be observed that the equations \eqref{AP2} and \eqref{ArP3} are related to $A_r$.
Next, substituting \eqref{Br} into \eqref{AP2} and \eqref{ArP3}, we have
   \begin{align}
&AP_2+P_2A^T_r-BB^T\left[ \begin{array}{cc}
            I & 0\\
            0 & Q_{322}^{-1}Q_{222}^T \\
          \end{array}
        \right]^T=0\label{jk1},\\
&A_rP_3+P_3A^T_r\nonumber\\
&~~~~~+\left[ \begin{array}{cc}
            I & 0\\
            0 & Q_{322}^{-1}Q_{222}^T \\
          \end{array}
        \right]BB^T\left[ \begin{array}{cc}
            I & 0\\
            0 & Q_{322}^{-1}Q_{222}^T \\
          \end{array}
        \right]^T=0.\label{ArP33}
\end{align}

From Eq.  \eqref{jk1}, we have
   \begin{align}
&\left[ \begin{array}{cc}
            I & 0\\
            0 & Q_{322}^{-1}Q_{222}^T \\
          \end{array}
        \right]AP_2+\left[ \begin{array}{cc}
            I & 0\\
            0 & Q_{322}^{-1}Q_{222}^T \\
          \end{array}
        \right]P_2A^T_r
        \nonumber\\&-\left[ \begin{array}{cc}
            I & 0\\
            0 & Q_{322}^{-1}Q_{222}^T \\
          \end{array}
        \right]BB^T\left[ \begin{array}{cc}
            I & 0\\
            0 & Q_{322}^{-1}Q_{222}^T \\
          \end{array}
        \right]^T=0,
\end{align}

combined with \eqref{ArP33} yields
   \begin{align}
&\left[ \begin{array}{cc}
            I & 0\\
            0 & Q_{322}^{-1}Q_{222}^T \\
          \end{array}
        \right]AP_2+\left[ \begin{array}{cc}
            I & 0\\
            0 & Q_{322}^{-1}Q_{222}^T \\
          \end{array}
        \right]P_2A^T_r
        \nonumber\\&+A_rP_3+P_3A_r^T=0. \label{deduceAr}
\end{align}

Due to Eq. \eqref{P2Q2}, it follows that
    \begin{align}
Q^{-1}_{322}Q^T_{22}P_{222}+P_{322}=0.
\end{align}

Using Eqs. \eqref{P212} and \eqref{P312}, we have
   \begin{align}
\tilde F_aP_{322}+Q^{-1}_{322}Q_{222}^T\bar F_a P_{222}=0. \label{deduceAr2}
\end{align}

Thus, it can be deduced from Eq. \eqref{deduceAr2} that
 $\tilde F_a=-Q_{322}^{-1}Q_{222}^T \bar F_a P_{222}P^{-1}_{322}$.
$\blacksquare$

\textbf{Remark 2.}
Theorem \ref{th3} provides necessary conditions for the design of $\tilde F_a$. However, since $P$, $Q$, $\hat A$, $\hat B$ and $\hat C$ are matrices that need to be calculated,
 the nonlinearity of equality constraints
 \eqref{barP} and \eqref{barQ} makes them difficult to solve. On the other hand, the coupling matrix $\tilde{F}_{pa}$ remains undetermined. To address these computational complexities, we
will transform the nonlinear equality constraints into a more tractable form and incorporate the coupling matrix into the solution.
%This strategy aims to enhance the solvability of the system while maintaining the integrity of the original constraints.

%In Theorem \ref{}, we present the system matrix of the ancillary system for the reduced augmented model. Since the model of the principal system remains unchanged before and after the reduction, \textcolor{red}{in the weak coupling scenario ($\bar F_{pa}=0$)}, the system matrix of the reduced system can be expressed as $A_r=\left[ \begin{array}{cc}
%            \tilde  G_{p} & 0\\
%            0 & -Q_{322}^{-1}Q_{222}^T \bar F_a P_{222}P^{-1}_{322} \\
%          \end{array}
%        \right]$, which can be interpreted as a general Markovian system. However, in general cases, the non-Markovian characteristics originate from the coupling with an environment exhibiting memory effects. Consequently, the coupling coefficient $\bar F_{pa}$ cannot be treated as zero, and we require another methods to design this coupling coefficient appropriately.
\subsection{Transformation of matrix equality constraints}
To make it more convenient for us to solve the optimization problem, in this section, we introduce a method to transform the nonlinear equality constraints into inequality constraints.
For the Lyapunov equation \eqref{barQ}, if we let \(  \hat{Q} = Q + \epsilon_Q \), where \( \epsilon_Q \) is a small positive definite matrix with appropriate dimensions, the equality constraint \eqref{barQ} becomes an inequality constraint
 \begin{align}
& \hat A^T {\hat Q} +  {\hat Q}  \hat A +   \hat C^T  \hat C
\nonumber\\&= \hat A^T { Q} +  { Q}  \hat A +   \hat C^T  \hat C+ \hat A^T\epsilon_Q+\epsilon_Q\hat A
\nonumber\\&=\hat A^T\epsilon_Q+\epsilon_Q\hat A< 0.
 \end{align}
 Consequently, we have \( \text{tr}(  \hat B^T \hat Q   \hat B) < \text{tr}(  \hat B^T  {Q}  \hat  B) \).
So we consider a positive real constant $\gamma$ as an upper bound of $||\Xi_G-\Xi_{G_r}||_{2}$; i.e., $||\Xi_G-\Xi_{G_r}||_{2}<\gamma$. Hence, (\ref{optp}) can be converted into the following optimization problem:

%For the stable linear quantum  system  \eqref{dx2} with physical realizability conditions \eqref{realr21} and \eqref{realr22},  find Hurwitz matrix $A_r$, real-valued matrices $B_r$, $C_r$ and positive definite matrices $  Q$ that minimize the square of the $\mathscr{H}_2$ norm \eqref{XiG} with  physical realizability conditions \eqref{realr21} and \eqref{realr22}; i.e,

\begin{align}
    & \min_{ {\hat Q}, A_r, B_r} \gamma^2 \label{opt2} \\
    \text{s.t.} \quad
    &  {\hat Q} = \begin{bmatrix}
         {\hat Q}_1 &  {\hat Q}_2 \\
         {\hat Q}_2^T &  {\hat Q}_3
    \end{bmatrix} > 0, \label{hatQ} \\
    &   \hat A^T  {\hat Q} +  {\hat Q}   \hat A +   \hat  C^T    \hat C < 0, \label{hatQAe} \\
    & \operatorname{tr}(   \hat B^T  {\hat Q}    \hat B) < \gamma^2, \label{hatQBe}\\
     \rm{Eqs.} &~ \eqref{realr21}-\eqref{pr3}.\nonumber
\end{align}

It can then be observed that once $\gamma^2$ is minimized, the $\mathscr{H}_2$ norm of the difference of the transfer functions between the original and the reduced model (\ref{XiG}) is also minimized.

 %Note that at this point, the Lyapunov equality \eqref{barQ} in the optimization problem \eqref{optp} has been transformed into inequality conditions \eqref{hatQAe}.
 % If we consider an equivalent transformation of the condition \eqref{hatQAe}  to obtain LMI conditions, we can derive equivalent conditions for \eqref{barQ}.
%  Firstly, system
%matrices $A$, $C$ in model \eqref{dx2} can be represented as block matrices $A=\left[\begin{array}{cc}A_{11}&A_{12}  \\A_{21} &A_{22}\\\end{array}\right]$,   $C=\left[\begin{array}{cc}C_1&C_{2}  \\\end{array}\right]$ .

Note that in the above derivation, when we solve the optimization problem we have not considered the matrices structure which results from the reduced augmented system (\ref{Ar1}), (\ref{Br1}), and (\ref{Cr1}). To take the matrices structure into account, on the one hand, we should preserve the principal system; i.e., \eqref{pr1} and \eqref{pr3} should be satisfied. On the other hand, we should also preserve the coupling structure. Concretely speaking, in the original system, the element $\bar F_{pa}$ of matrix $A$ represents the coupling strength between the principal oscillator and the ancillary system.

The coupling matrix  must strictly preserve its tensor product structure, as defined in Eq. \eqref{real25}, to ensure that the coupling strength remains rigorously real-valued.
By integrating the structure of system matrices \( B_r \) and \( \tilde F_a \) from Theorems \ref{th4} and \ref{th3} with the configuration of coupling matrix \( \bar F_{pa} \) in \eqref{real25}, we derive the following theorem.

% Denote n-dimensional vector \(S_n=[1,0,...,0] \) and r-dimensional vector \(S_r=[1,0,...,0] \), according to \eqref{LE}, we have $\bar F_{pa}=\frac{\sqrt{\gamma_p\kappa_p}}{2}K_n$, where $K_n=\mathcal{M}(S_n)$. Define \( K_r =\mathcal{M}(S_r)\),
% the main findings are as follows.

\begin{theorem}\label{th5}For the original augmented system model of a linear non-Markovian quantum system \eqref{dx2}, if there exist  positive scalars $\gamma$, $\alpha_1$ and $\alpha_2$, positive block diagonal symmetric  matrices $  \hat Q_{1}\in\mathbb{R}^{2n\times2n}$,
    $ \hat  Q_{3}\in\mathbb{R}^{2r\times2r}$,
   $  M\in\mathbb{R}^{2n\times2n}$ and matrices $  \hat Q_{2}\in\mathbb{R}^{2n\times2r}$, $\beta\in\mathbb{R}^{m\times r}$ such that the following optimization problem is solvable
    \begin{align}
& \min_{ {\hat Q}_1, {\hat Q}_2,  {\hat Q}_3, M,\beta} \gamma^2 \label{opt3} \\
   {\rm s.t.} \quad
     &\hat Q=\begin{bmatrix}
\begin{array}{cc}
\hat Q_1 & \hat Q_2 \\
\hat Q_2^T & \hat Q_3 \\
\end{array}
\end{bmatrix}>0,\label{th21}\\
    & \left[\begin{array}{c}
\alpha_1A^T  \hat Q_1+ \alpha_1 \hat Q_1A+\alpha_1 C^TC \\
 \# \\
\end{array}\right.\nonumber\\
&\quad\ \left.\begin{array}{c}
MA\\
A^T \hat  Q_1+  \hat Q_1A  \\
\end{array}\right]<0,\label{36}
\\
&\left[\begin{array}{c}
\alpha_2 A^T  \hat Q_1+ \alpha_2 \hat  Q_1A+\alpha_2C^TC \\
 \# \\
\end{array}\right.\nonumber\\
&\quad\ \left.\begin{array}{c}
A^T\hat Q_2 +\hat Q_2A_{r1} -C^TC_{r}   \\
A_{r1} ^T \hat Q_3+ \hat Q_3 A_{r1}  +C_{r}^TC_{r} \\
\end{array}\right]<0\label{37},\\
%
%
%&\left[\begin{array}{cc}\alpha_2 A^T  \hat Q_1+ \alpha_2 \hat  Q_1A+\alpha_2C^TC&  A^T\hat Q_2 +\hat Q_2A_{r1} -C^TC_{r}\\ \# &A_{r1} ^T \hat Q_3+ \hat Q_3 A_{r1}  +C_{r}^TC_{r} \\\end{array}\right]\nonumber\\&<0\label{37},\\
%   \end{align}
%   \begin{align}
&\alpha_1+\alpha_2=1,\label{41}
%\end{align}
%\begin{align}
\\
%&\mathrm{tr}\left(\bar EB^T(3M+\hat Q_1)B\bar E+2B_{r1}^T\hat Q_2^TB+B_{r1}\hat Q_3B_{r1}\right)\nonumber\\&<\gamma^2,
&\mathrm{tr}\left(\left[\begin{array}{cc}B^T&B^T_r \\\end{array}\right]\left[\begin{array}{cc}  \hat Q_1&  \hat Q_2  \\  \hat Q_2^T &  \hat Q_3\\\end{array}\right]\left[\begin{array}{c}B\\B_r \\\end{array}\right]\right)<\gamma^2,
\label{38}
\\
&   \hat Q_1  \hat Q_2-  \hat Q_2  \hat Q_3=0,\label{39}\\
&M-  \hat Q_2 \bar E \hat  Q_3 ^{-1} \hat  Q_2^T=0, \label{40}
\end{align}
   then, the reduced augmented system model of the linear non-Markovian quantum system \eqref{dxr} is given as
 \begin{align}
  A_r&=  \begin{bmatrix}
\begin{array}{cc}
 \bar F_{p} & \beta\otimes I \\
-(\beta\otimes I)^T & ETAVE^T \\
\end{array}
\end{bmatrix},\nonumber\\
   B_r&=  \begin{bmatrix}
\begin{array}{cc}
                                  \bar G_{p} & 0_{2m\times2r} \\
                                   0_{2r\times2m} & ETBE^T \\
\end{array}
\end{bmatrix},  \label{th5Ar}
\end{align}
 where  $  T= \hat Q_3^{-1}  \hat Q_2^{T} $, $  V=  \hat Q_1^{-1} \hat  Q_2$,
 $A_{r1} =\begin{bmatrix}
\begin{array}{cc}
\bar F_p &  \beta\otimes I  \\
 -(\beta \otimes I)^T & 0_{2r\times2r} \\
\end{array}
\end{bmatrix}$,
 $
   B_{r1}=\begin{bmatrix}
\begin{array}{cc}
                                    \bar G_{p} & 0_{2m\times2r} \\
                                   0_{2r\times2m} & 0_{2r\times2r} \\
\end{array}
\end{bmatrix}$,  $E=\begin{bmatrix}
\begin{array}{cc}
0_{2r\times 2m} & I _{2r\times 2r}
\end{array}
\end{bmatrix}$, $
   \bar E=\begin{bmatrix}
\begin{array}{cc}
                                   0_{2m\times 2m} & 0_{2m\times 2r}\\
                                   0_{2r\times 2m} & I _{2r\times 2r}\\
\end{array}
\end{bmatrix}$.
\end{theorem}

 \textbf{Proof.}
% In this proof, we prove the equivalence between problems \eqref{opt2} and \eqref{opt3} under condition \eqref{th5Ar}. Specifically, we transform the nonlinear matrix inequality conditions \eqref{hatQAe} and \eqref{hatQBe} into their corresponding linear matrix inequality counterparts, namely \eqref{36} - \eqref{41} and \eqref{38} respectively. Furthermore, we show that the matrices \eqref{th5Ar} for the reduced model remain physically realizable.
Firstly, we pre-multiply and post-multiply $\left[\begin{array}{cc}I&0 \\ 0&  \bar E\hat Q_2^T\hat Q_1^{-1}\\\end{array}\right]$ and $\left[\begin{array}{cc}I&0 \\ 0&  \hat Q_1^{-1}\hat Q_2\bar E\\\end{array}\right]$ on both sides of \eqref{36},  respectively, which yields

\begin{align}
&\left[\begin{array}{cc}I&0 \\ 0&  \bar E\hat Q_2^T\hat Q_1^{-1}\\\end{array}\right]\nonumber \left[\begin{array}{c}
\alpha_1A^T  \hat Q_1+ \alpha_1 \hat Q_1A+\alpha_1C^TC \\
 \# \\
\end{array}\right.\\
&\quad\ \left.\begin{array}{c}
MA   \\
A^T  \hat Q_1+  \hat Q_1A  \\
\end{array}\right]\left[\begin{array}{cc}I&0 \\ 0&  \hat Q_1^{-1}\hat Q_2\bar E\\\end{array}\right]\nonumber\\%<0
&=
%\left[\begin{array}{c}
%\alpha_1A^T\hat Q_1+\alpha_1\hat Q_1A+\alpha_1C^TC \\
% \# \\
%\end{array}\right.\nonumber\\
%&\quad\ \left.\begin{array}{c}
%\hat Q_2A_{r2}   \\
%A_{r2}^T\hat Q_3+ \hat Q_3A_{r2} \\
%\end{array}\right]
%<0,\label{6121}
\left[\begin{array}{cc}
\alpha_1A^T\hat Q_1+\alpha_1\hat Q_1A+\alpha_1C^TC & \hat Q_2A_{r2}\\
 \# & A_{r2}^T\hat Q_3+ \hat Q_3A_{r2}
\end{array}\right]<0,\label{6121}
\end{align}
where
\begin{align}
A_{r2}&=  \begin{bmatrix}
\begin{array}{cc}
0_{2m\times2m} & 0_{2m\times2r} \\
0_{2r\times2m} & ETAVE^T \\
\end{array}
\end{bmatrix}.
\end{align}
Combining (\ref{th5Ar}) with Eqs. \eqref{37} and \eqref{41}, we obtain
\begin{align}
&
\left[\begin{array}{c}
\alpha_1A^T\hat Q_1+\alpha_1\hat Q_1A+\alpha_1C^TC \\
 \# \\
\end{array}\right.\nonumber\\
&\quad\ \left.\begin{array}{c}
\hat Q_2A_{r2}    \\
A_{r2}^T\hat Q_3+ \hat Q_3A_{r2} \\
\end{array}\right]+\nonumber\\
&\left[\begin{array}{cc}\alpha_2 A^T  \hat Q_1+ \alpha_2 \hat  Q_1A+\alpha_2C^TC&  A^T\hat Q_2 +\hat Q_2A_{r1} -C^TC_{r}\\ \# &A_{r1} ^T \hat Q_3+ \hat Q_3 A_{r1}  +C_{r}^TC_{r} \\\end{array}\right]
\nonumber\\
 &=\left[\begin{array}{cc}A^T \hat Q_1+ \hat Q_1A+C^TC&A^T \hat Q_2+\hat Q_2A_r-C^TC_r \\ \# &A_r^T\hat Q_3+ \hat Q_3A_r+C_r^TC_r\\\end{array}\right]\nonumber\\
 &=\hat A^T  {\hat Q} +  {\hat Q}   \hat A +   \hat  C^T    \hat C<0.
\end{align}
Hence, we have shown that \eqref{hatQAe} is equivalent to the LMIs \eqref{36}-\eqref{41}.

%Next, we show the equivalence of \eqref{hatQBe} to \eqref{38}.
%To eliminate the nonlinear term $ \hat B^T  {\hat Q}  \hat  B$ in \eqref{hatQBe}, we employ condition \eqref{th21} and obtain
%\begin{align}\label{BeQ}
%&\mathrm{tr}( \hat B^T  \hat Q \hat B)
%=\mathrm{tr}\left(\left[\begin{array}{cc}B^T&B^T_r \\\end{array}\right]\left[\begin{array}{cc}  \hat Q_1&  \hat Q_2  \\  \hat Q_2^T &  \hat Q_3\\\end{array}\right]\left[\begin{array}{c}B\\B_r \\\end{array}\right]\right)\nonumber\\
%&=\mathrm{tr}\left( B^TQ_1B+B_r^TQ_2^TB+B^TQ_2B_r+B_r^TQ_3B_r \right)\nonumber\\
%&=\mathrm{tr}\left(\bar EB^T(3M+\hat Q_1)B\bar E+2B_{r1}^T\hat Q_2^TB+B_{r1}\hat Q_3B_{r1}\right).
%\end{align}
%This transformation converts \eqref{hatQBe} into the condition \eqref{38}.

Furthermore, by employing \eqref{39}, we derive
%\begin{align}
%&EV^TA^TT^TE^T+ETAVE^T+TBE^TEB^TT^T \nonumber\\
%&=EV^TA^TVE^T+EV^TAVE^T+V^TBB^TV
%\nonumber\\
%&=0
%\end{align}
%and
%the physical realization conditions as follows
\begin{align}
&A_r+A_r^T+B_rB_r^T\nonumber\\&= \begin{bmatrix}
\begin{array}{cc}
\bar F_p+\bar F_p^T+\bar G_p\bar G_p^T & \beta\otimes I-\beta\otimes I \\
(\beta\otimes I)^T -(\beta\otimes I)^T& 0_{2r\times2r} \\
\end{array}
\end{bmatrix}\\\nonumber &
=0.
\end{align}
From the structure of \eqref{th5Ar}, it is evident that the matrix structure of the augmented system model of the linear non-Markovian quantum system and the matrices of the principal system are preserved.
 $\blacksquare$

In Theorem \ref{th5}, the elements $\beta_{ij}$ of matrix $\beta$ correspond to the elements of the coupling matrix $\tilde F_{pa}$.
Although Eqs. \eqref{th21}, \eqref{36} and \eqref{41} have been transformed to LMI conditions in
 Theorem \ref{th5}, Eqs. \eqref{37}, \eqref{39} and \eqref{40} are still nonlinear matrix equations, presenting significant computational challenges. To address this difficulty, we will employ a lifting variables approach to transform the nonlinear terms to linear terms.

An alternative formulation for our optimization problem could involve replacing the objective function with ${\rm tr}(\hat{C}P\hat{C}^T)$, the optimization variable with $P$, and the constraint \eqref{barQ} with \eqref{barP}. However, since these two formulations are equivalent, we will not explore it in this work.

\subsection{Matrix Lifting Method for Model Reduction}

In this section, we use a matrix lifting method to transform  Eqs. \eqref{39} and \eqref{40} into linear forms.
This transformation requires the introduction of appropriate matrix lifting variables and their associated equation constraints \cite{r8}.

 Firstly, six matrix lifting variables $  W_{1-6}$ are given as: $  W_1=  \hat Q_1 $, $  W_2=   \hat Q_2 $, $  W_3=  W_1   \hat Q_2$, $  W_4=  W_2   \hat Q_3$, $  W_5=   \hat Q_3^{-1}$ and $  W_6=   \hat Q_2^T$.
 Now, we define a matrix $  {\mathbf Z}$ be a symmetric matrix.
  By replacing $   \hat Q_1$, $   \hat Q_2$, $   \hat Q_3$, $  M$, $\hat Q_3^{-1}$, $   \hat Q_2^T$, $\beta$, $\beta^T$, $\hat Q_{211}$, $\hat Q_{222}$, $\hat Q_{311}$, $\hat Q_{322}$
  with $  Z_{x_1,1}$, $  Z_{x_2,1}$, $  Z_{x_3,1}$, $  Z_{x_4,1}$, $  Z_{x_5,1}$, $  Z_{x_6,1}$,  $  Z_{x_7,1}$, $Z_{x_8,1}$, $Z_{x_9,1}$, $Z_{x_{10},1}$, $Z_{x_{11},1}$, $Z_{x_{12},1}$,   respectively, where
  \begin{align}
  Z_{i,j}=[  {\mathbf Z}_{kl}].
\end{align}
The matrix $\mathbf Z$ satisfies the following constraints:
\begin{align}
 &  \textbf Z\geq0,\label{81}\\
&  Z_{0,0}-I_n =0,\label{148}\\
  &  Z_{x_1,1}-  Z_{1,x_1}=0 ,\\
  &  Z_{x_3,1}-  Z_{1,x_3}=0,\\
  &  Z_{x_4,1}-  Z_{1,x_4}=0 ,\\
  &  Z_{v_1,1}-  Z_{x_2,x_5}=0,\\
    &  Z_{v_2,1}-  Z_{x_5,x_6}=0,\label{153}\\
    &  Z_{v_3,1}-  Z_{x_1,x_2}=0,\label{155}\\
    &   Z_{v_4,1}-  Z_{x_2,x_3}=0, \label{156}\\
    & Z_{v_5,1}-Z_{x_{9},1}(Z_{x_7,1}\otimes I) =0, \\
    & Z_{v_6,1}-Z_{x_{10},1}(Z_{x_8,1}\otimes I)=0, \\
    & Z_{v_7,1}-Z_{x_{11},1}(Z_{x_7,1}\otimes I)=0, \\
    & Z_{v_8,1}-Z_{x_{12},1}(Z_{x_8,1}\otimes I)=0,
\end{align}
where
\begin{align}
  Z_{a,b}=  Z_{a,1}  Z_{b,1},\label{zab2}
\end{align}
 with $a,b\in\{v_1,..., v_8\}\cup\{x_1,...,x_{12}\}$.
Then, the scalar variables to be evaluated can be written as $\alpha_1$, $\alpha_2$ and $\beta$.
With these,  constraints (\ref{th21})-(\ref{38}) can be rewritten as
 \begin{align}
   &   Z_{x_1,1},  Z_{x_3,1}>0, \label{119} \\
     & \left[\begin{array}{c}
\alpha_1A^T  Z_{x_1,1}+ \alpha_1 Z_{x_1,1}A+\alpha_1C^TC \\
 \# \\
\end{array}\right.\nonumber\\
&\quad\ \left.\begin{array}{c}
Z_{x_4,1}A   \\
A^T  Z_{x_1,1}+  Z_{x_1,1}A  \\
\end{array}\right]<0,\label{120}
\\
&\left[\begin{array}{cc}\alpha_2 A^T  Z_{x_1,1}+ \alpha_2  Z_{x_1,1}A+\alpha_2C^TC& A^TZ_{x_2,1} +\Pi_1\\ \# & \Pi_2^T+ \Pi_2+C_{r}^TC_{r} \\\end{array}\right]<0,\\
&
\mathrm{tr}\left(\left[\begin{array}{cc}B^T&\Pi_3^T \\\end{array}\right]\left[\begin{array}{cc}  \hat Q_1&  \hat Q_2  \\  \hat Q_2^T &  \hat Q_3\\\end{array}\right]\left[\begin{array}{c}B\\ \Pi_3 \\\end{array}\right]\right)<\gamma^2,
\\
&\alpha_1>0, \alpha_2>0,\label{123} \\
&\alpha_1+\alpha_2=1,\label{124}
\end{align}
where $\Pi_1=\begin{bmatrix}
         -Z_{x_9,1}\bar F_p &  Z_{v_5,1} \\
         -Z_{v_6,1} &  0
    \end{bmatrix}-C^TC_{r}$, $\Pi_2=\begin{bmatrix}
         -Z_{x_{11},1}\bar F_p &  Z_{v_{7},1} \\
         -Z_{v_{8},1} &  0
    \end{bmatrix}$, $\Pi_3=\begin{bmatrix}
\begin{array}{cc}
                                  \bar G_{p} & 0_{2m\times2r} \\
                                   0_{2r\times2m} & EZ_{v_2,1}BE^T \\
\end{array}
\end{bmatrix}$.
Next, we consider the transformation of the nonlinear equality matrix constraint \eqref{39} into linear matrix equality constraints.
One can easily obtain
  \begin{align}
Z_{v_3,1}=Z_{v_1,x_2}=Z_{v_1,1}Z_{x_2,1}=Z_{x_1,1}Z_{x_2,1}= \hat Q_1 \hat Q_2,\\
Z_{v_4,1}=Z_{v_2,x_3}=Z_{v_2,1}Z_{x_3,1}=Z_{x_2,1}Z_{x_3,1}= \hat Q_2 \hat Q_3,
\end{align}
 and the physical realizability constraint \eqref{39} can be replaced by
\begin{align}
  Z_{v_3,1}-Z_{v_4,1}=0,
\end{align}
Similarly,  the nonlinear equality matrix constraint \eqref{40} can be transformed into linear matrix equality constraints
\begin{align}
      &I_r-  Z_{x_3,x_5}=0\label{154}\\
    & Z_{v5,1}-Z_{x2,x5}=0,\\
    & Z_{v6,1}-Z_{v5,x6}=0,\\
    & Z_{x4,1}-Z_{v6,1}=0.\label{160}
\end{align}

This leads us to reformulate the optimization problem \eqref{opt3}.

For the non-Markovian oscillator  \eqref{dx2} with physical realizability conditions \eqref{real21}-\eqref{real25},  find a matrix $  {\mathbf Z}$ that minimizes the square of the $\mathscr{H}_2$ norm \eqref{XiG}; i.e,
\begin{align}
    & \min_{ {\mathbf Z},\alpha_1,\alpha_2} \gamma^2 \label{opt60}\\
    \text{s.t.} \quad &\text{Eqs.}\ \eqref{81}-\eqref{160}.\nonumber
\end{align}

As outlined in Ref. \cite{r8}, it is important to select a reasonable initial point for the algorithm, we suggest solving LMI constraints \eqref{119}, \eqref{120}, \eqref{123} and \eqref{124} to obtain the heuristic starting point.
Thus, the optimization problem \eqref{opt60} is solved by SeDumi \cite{sedumi} and Yalmip \cite{yalmip}.

\section{Numerical Example of a non-Markovian Quantum Oscillator System}\label{sec5}
In this section, we consider reducing the dimension of an augmented system model for a non-Markovian quantum oscillator system, where the principal system consists of two harmonic oscillators and the ancillary system consists of three harmonic oscillators. The main parameters of the system are given as follows. The frequencies of the principal and  ancillary oscillators are $\omega_1=10.85{\rm GHz}$, $\omega_2=9.74{\rm GHz}$,  $\omega_3=10.03{\rm GHz}$, $\omega_4=8.93{\rm GHz}$, $\omega_5=5.06{\rm GHz}$, respectively. The damping rates of the principal and ancillary systems are $\gamma_1= 0.954{\rm GHz}$, $\gamma_2= 0.987{\rm GHz}$, $\gamma_3= 0.848{\rm GHz}$, $\gamma_4=  1.034{\rm GHz}$, $\gamma_5= 0.775{\rm GHz}$, respectively. The direct coupling strengths are  $\kappa_1=1.25{\rm GHz}$, $\kappa_2=1.14{\rm GHz}$, respectively.
Thus, the original augmented system matrices are given as

\begin{eqnarray}
   \bar F_{p}&=&{\rm diag}\left(\left[
                                \begin{array}{cc}
                                 -\frac{\gamma_{1}}{2}  & \omega_1\\
             -\omega_{1} & -\frac{\gamma_{1}}{2}\\
                                \end{array}
                              \right],\left[
                                        \begin{array}{cc}
                                          -\frac{\gamma_2}{2} &  \omega_2\\
               -\omega_2  &-\frac{\gamma_2}{2}\\
                        \end{array}
                                      \right]
   \right), \label{exfp}\\
%\end{equation}
%\begin{align}
%  \bar F_{p}&=\left[ \begin{array}{cccc}
%            -\frac{\gamma_{1}}{2}  & \omega_1 & 0 &0\\
%             -\omega_{1} & -\frac{\gamma_{1}}{2} &  0 &0 \\
%              0& 0 &-\frac{\gamma_2}{2} &  \omega_2\\
%               0 &0 &-\omega_2  &-\frac{\gamma_2}{2}
%          \end{array}
%        \right],
%                \end{align}
%\begin{eqnarray}
   \bar F_{a}&=&{\rm diag}{\bigg(}\left[
                                \begin{array}{cc}
                                -\frac{\gamma_{3}}{2}  & \omega_3 \\
             -\omega_{3} & -\frac{\gamma_{3}}{2} \\
                                \end{array}
                              \right],\left[
                                        \begin{array}{cc}
                                          -\frac{\gamma_4}{2} &  \omega_4\\
              -\omega_4  &-\frac{\gamma_4}{2}\\
                                        \end{array}
                                      \right],\nonumber\\
   &&~~~~~~~~~~~~~~~~~~~~~~ \left[
                                        \begin{array}{cc}
                                          -\frac{\gamma_5}{2} &  \omega_5\\
             -\omega_5  &-\frac{\gamma_5}{2}\\
                                        \end{array}
                                      \right]
   {\bigg)}\label{exfa}\\
      \bar F_{pa}&=&\left[
                 \begin{array}{ccc}
                   \frac{\sqrt{\kappa_1\gamma_3}}{2} & \frac{\sqrt{\kappa_1\gamma_4}}{2} & \frac{\sqrt{\kappa_1\gamma_5}}{2}  \\
                   \frac{\sqrt{\kappa_2\gamma_3}}{2} & \frac{\sqrt{\kappa_2\gamma_4}}{2}  &  \frac{\sqrt{\kappa_2\gamma_5}}{2} \\
                 \end{array}
               \right]\otimes I_2,\label{exfpa}\\
  \bar G_p&=&{\rm diag}\left (\sqrt{\gamma_1}, \sqrt{\gamma_1}, \sqrt{\gamma_2},\sqrt{\gamma_2}\right),\label{exgp}\\
     \bar G_a&=&{\rm diag}\left (\sqrt{\gamma_3}, \sqrt{\gamma_3}, \sqrt{\gamma_4},\sqrt{\gamma_4}, \sqrt{\gamma_5},\sqrt{\gamma_5}\right).\label{exga}
%     D&=&I
\end{eqnarray}

Here, it is crucial to emphasize that existing model reduction techniques, such as the singular perturbation method \cite{r4}, the quasi-balanced truncation method \cite{r6} and other methods, are not applicable in this example because we must ensure the preservation of the primary system in the reduced model.

In the quadrature representation, we reduce the number of the  harmonic oscillators in the ancillary system model from 3 to 2 according to $\mathscr{H}_2$ norm performance, which also satisfies the physical realizable conditions.
Based on Theorem  \ref{th5}, we  derive the reduced model matrices
as follows:
 \begin{align}\tilde  F_{a}&=\left[ \begin{array}{cc}
            -0.6278  & 10.0793\\
    -10.0696 &  -0.6277
          \end{array}
        \right],
                \end{align}
        \begin{align}
       \tilde  F_{pa}&=\left[ \begin{array}{cc}
 0.5528    &     0\\
      0   & 0.5528\\
 0.5262    &     0\\
       0  &  0.5262
          \end{array}
        \right],
                \end{align}
        \begin{align}
\tilde  N_{a}&=\left[ \begin{array}{cc}
 1.1201 &  -0.0310\\
 0.0224  &  1.1202
          \end{array}
        \right].
        \end{align}
%$A_r = \left[\begin{array}{cccc}
%-0.5724 & 10.3500 & 1.8869 & 0 \\
%-10.3500 & -0.5724 & 0 & 1.8869 \\
%-1.8869 & 0 & -0.5322 & -9.0459 \\
%0 & -1.8869 & 9.0327 & -0.5930\\
%\end{array}\right],
%$
%
%$B_r = \left[\begin{array}{cccc}
%1.0700 & 0 & 0 & 0  \\
%0 & 1.0700 & 0 & 0 \\
%0 & 0 & -1.0073 & 0.2233  \\
%0 & 0 & 0.2232 & 1.0659 \\
%\end{array}\right],
%$
%
%$C_r = \left[\begin{array}{cccc}
%-1.0700 & 0 & 0 & 0 \\
%0 & -1.0700 & 0 & 0 \\
%0 & 0 & 1.0073 & -0.2232 \\
%0 & 0 & -0.2233 & -1.0659 \\
%\end{array}\right].
%$
It is easy to check that the reduced model satisfies physical realizability, as given in \eqref{realr21}, \eqref{realr22} and \eqref{realr23}. In addition, the reduced model preserves the essential properties of the harmonic oscillators of the principal system and reduces  the two harmonic oscillators to one in the ancillary system .

\begin{figure}[htbp]
\centerline{\includegraphics[width=9cm]{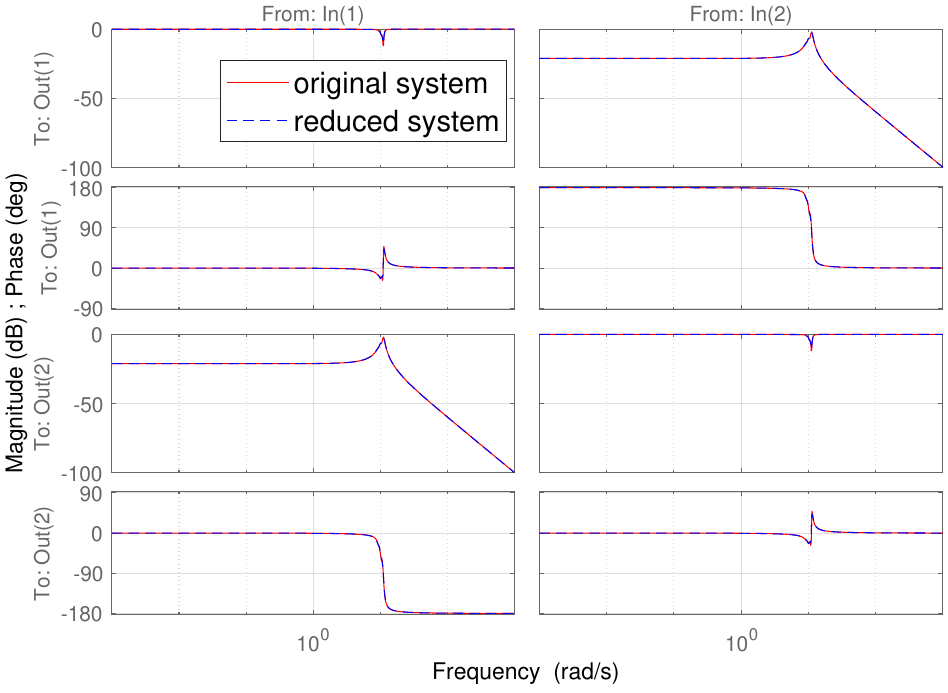}}
\caption{Comparison of frequency responses between the original and reduced models for Input 1, 2 and Output 1, 2}
\label{fig1}
\end{figure}

\begin{figure}[htbp]
\centerline{\includegraphics[width=9cm]{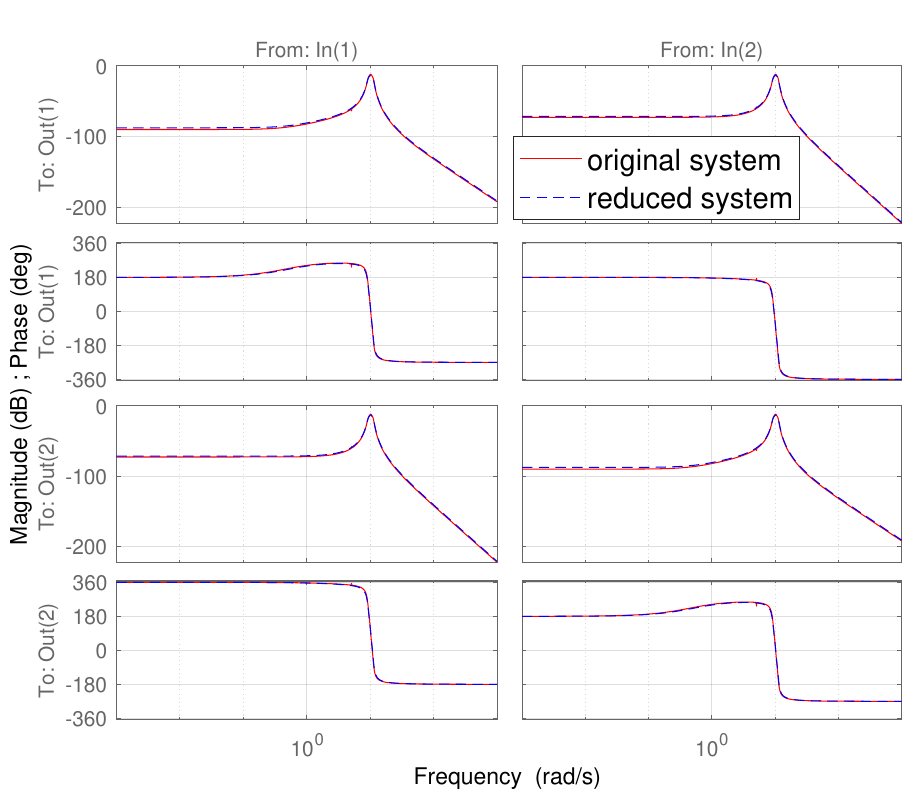}}
\caption{Comparison of frequency responses between the original and reduced models for Input 1, 2 and Output 3, 4}
\label{fig2}
\end{figure}
After calculation, we obtain $||\Xi_G-\Xi_{G_r}||=0.1746$.
Figs. \ref{fig1} and \ref{fig2} present a comparative analysis of the input-output response between the original and reduced models through their Bode plots. Notably, the amplitude-frequency characteristics of the reduced model closely match those of the original system across the entire frequency range, demonstrating remarkable fidelity in magnitude response. Furthermore, the phase-frequency characteristics of the two models exhibit significant alignment, particularly in the low-frequency region, where the phase responses are nearly indistinguishable. This close correspondence can be attributed to the preservation of the principal system's dynamics and careful optimization of the coupling matrices during the reduction process.

 %Such modifications are particularly valuable for optimizing system performance and facilitating theoretical analysis in studies involving time-delay effects in quantum systems.
% According to Fig. \ref{fig1}, there is a noticeable jump in the curve around the frequency of $10$, which is attributed to resonance occurring near the system's natural frequency. Since the reduced model retains most of the high-frequency features, its response curve matches the original model well at higher frequencies.
%In Fig. \ref{fig1.01}, we illustrate the decreasing trend of the objective function $\gamma$ in Theorem 6 during the optimization process, ensuring that the results we obtain are indeed the optimal solution.
 %  Overall, the model reduction appears to have been performed with high accuracy, ensuring that the simplified model remains a reliable representation of the original system.

%$A = \begin{pmatrix}
%-0.5724 & 10.3500 & 0.6171 & 0 & 0 & 0 \\
%-10.3500 & -0.5724 & 0 & 0.6171 & 0 & 0 \\
%-0.6171 & 0 & -0.5922 & 9.4400 & 0.6171 & 0 \\
%0 & -0.6171 & -9.4400 & -0.5922 & 0 & 0.6171 \\
%0 & 0 & -0.6171 & 0 & 0 & 1.0300 \\
%0 & 0 & 0 & -0.6171 & -1.0300 & 0
%\end{pmatrix}$,
%$
%B = \begin{pmatrix}
%1.0700 & 0 & 0 & 0 & 0 & 0 \\
%0 & 1.0700 & 0 & 0 & 0 & 0 \\
%0 & 0 & 1.0883 & 0 & 0 & 0 \\
%0 & 0 & 0 & 1.0883 & 0 & 0 \\
%0 & 0 & 0 & 0 & 0 & 0 \\
%0 & 0 & 0 & 0 & 0 & 0
%\end{pmatrix}$,
%$
%C = \begin{pmatrix}
%-1.0700 & 0 & 0 & 0 & 0 & 0 \\
%0 & -1.0700 & 0 & 0 & 0 & 0 \\
%0 & 0 & -1.0883 & 0 & 0 & 0 \\
%0 & 0 & 0 & -1.0883 & 0 & 0 \\
%0 & 0 & 0 & 0 & 0 & 0 \\
%0 & 0 & 0 & 0 & 0 & 0
%\end{pmatrix}.
%$

\section{Conclusion}\label{sec6}
%This paper proposes an optimization strategy for model reduction based on the $\mathscr{H}_2$ norm, ensuring a physically realizable reduced model that closely approximates the original system. The method minimizes the $\mathscr{H}_2$ norm difference, enabling efficient application of the augmented system model to real-time feedback control in non-Markovian quantum systems. Future work could explore extending this approach to optimize $\mathscr{H}_\infty$ performance, enhancing robustness and performance in quantum systems.

This paper presents a novel $\mathscr{H}_2$-based model reduction approach to reducing the dimension of augmented models for non-Markovian quantum systems.
The proposed method has formulated the reduction as a constrained optimization problem that ensures physical realizability under non-Markovian dynamics, with three key advances: first, establishing rigorous necessary and sufficient conditions for physical realizability that are stricter than Markovian requirements; second, deriving input matrix and ancillary system matrix constraints for the reduced model; third, developing a structure preserving design method for the reduced augmented system.
 Compared to model reduction methods in Markovian systems, the proposed method effectively preserves the non-Markovian characteristics of the original system.
Simulations have confirmed its effectiveness.
Future work could explore extending this approach to optimize $\mathscr{H}_\infty$ performance, enhancing robustness and performance in quantum systems.

\vspace{12pt}
%\color{red}
%IEEE conference templates contain guidance text for composing and formatting conference papers. Please ensure that all template text is removed from your conference paper prior to submission to the conference. Failure to remove the template text from your paper may result in your paper not being published.


\begin{thebibliography}{00}

\bibitem{bg1}
D.  d'Alessandro. (2021). Introduction to quantum control and dynamics. Chapman and hall/CRC.

\bibitem{bg2}
Z. Zhou, R. Sitler, Y. Oda, K. Schultz, G. Quiroz. (2023). Quantum crosstalk robust quantum control. \textit{Phys. Rev. Lett}, 131(21), 210802.

\bibitem{bg3}
D. Dong, X. Xing, H. Ma, C. Chen, Z. Liu, H. Rabitz. (2019). Learning-based quantum robust control: algorithm, applications, and experiments. \textit{IEEE T. Cybern}, 50(8), 3581-3593.





%\bibitem{r3} D. E. Rivera, M. Morari. (1990). Low-order SISO controller tuning methods for the $H_2$, $H_\infty$ and $\mu$ objective functions. \textit{Automatica}, 26(2), 361-369.
%\bibitem{r2} M. R. James, H. I. Nurdin, I. R. Petersen. (2008). $H_\infty$ control of linear quantum systems. \textit{IEEE Trans. Autom. Control}, 53(8), 1787-1803.
\bibitem{kumar2014}
A. Kumar, M. Sarovar. (2014). On model reduction for quantum dynamics: symmetries and invariant subspaces. \textit{J. Phys. A: Math. Theor.}, 48(1), 015301.

%\bibitem{re1} L. Bouten, R. van Handel, and A. Silberfarb. (2008). Approximation and limit  theorems for quantum stochastic models with unbounded coefficients,  \textit{J. Functional Anal}., 254, 3123¨C3147.
%\bibitem{re2} H. I. Nurdin and J. E. Gough. (2012). On structure preserving transformations  of the Ito generator matrix for model reduction of quantum feedback  networks, \textit{Phil. Trans. R. Soc. A}, 370, 5422¨C5436.


\bibitem{r4} I. R. Petersen. (2010). Singular perturbation approximations for a class of linear complex quantum systems. \textit{Proc. Amer. Control Conf.}, Baltimore, MD, 1898-1903.
\bibitem{r5} I. R. Petersen. (2012). Low frequency approximation for a class of linear quantum systems using cascade cavity realization. \textit{Syst. Control Lett.}, 6(1), 173-179.

\bibitem{r6} H. I. Nurdin. (2014). Structures and transformations for model reduction of linear quantum systems. \textit{IEEE Trans. Autom. Control}, 59(9), 2413-2425.
\bibitem{r7} O. Techakesari, H. I. Nurdin. (2016). Tangential interpolatory projection for model reduction of linear quantum systems. \textit{IEEE Trans. Autom. Control}, 62(1), 5-17.

\bibitem{chase2009}
B. A. Chase, and J. M. Geremia. (2008). Collective processes of an ensemble of
spin-1/2 particles, \textit{Phys. Rev. A}
78(5), 052101.

\bibitem{ticozzi2023}
G. Grigoletto, F. Ticozzi. (2023). Model reduction for quantum systems: Discrete-time quantum walks and open Markov dynamics. \textit{arXiv preprint arXiv:2307.06319}.

\bibitem{ticozzi2024b}
G. Grigoletto, Y. Tao, F. Ticozzi, L. Viola. (2024). Exact Model Reduction for Continuous-Time Open Quantum Dynamics. \textit{arXiv preprint arXiv:2412.05102}.

   \bibitem{ticozzi2024}
G. Grigoletto, F. Ticozzi. (2024). Exact model reduction for discrete-time conditional quantum dynamics. \textit{IEEE Control Systems Letters}.



\bibitem{ticozzi2025}
G. Grigoletto, C. Pellegrini, F. Ticozzi. (2025). Quantum model reduction for continuous-time quantum filters. \textit{arXiv preprint arXiv:2501.13885}.

%\bibitem{elliott2020}
%T. J. Elliott, C. Yang, F. C. Binder, A. J. Garner, J. Thompson, M. Gu. (2020). Extreme dimensionality reduction with quantum modeling. \textit{Phys. Rev. Lett}, 125(26), 260501.
%
%\bibitem{elliott2021}
%T. J. Elliott. (2021). Quantum coarse graining for extreme dimension reduction in modeling stochastic temporal dynamics. \textit{PRX Quantum}, 2(2), 020342.



\bibitem{fan2024}
W. Fan, H. E. Tureci. (2024). Model Order Reduction for Open Quantum Systems Based on Measurement-adapted Time-coarse Graining. \textit{arXiv preprint arXiv:2410.23116}.





%\bibitem{chen2015}
%Q. Chen, J. Li, C. Yam, Y. Zhang, N. Wong, G. Chen. (2015). An approximate framework for quantum transport calculation with model order reduction. \textit{J. Comput. Phys.}, 286, 49-61.






%\bibitem{nm1} C. Li, G. Guo, J. Piilo. (2020). Non-Markovian quantum dynamics: What is it good for? \textit{EPL}, 128(3), 30001.



%\bibitem{nm3} S. Cialdi, C. Benedetti, D. Tamascelli, S. Olivares, M. G. Paris, B. Vacchini. (2019). Experimental investigation of the effect of classical noise on quantum non-Markovian dynamics. \textit{Phys. Rev. A}, 100(5), 052104.

%\bibitem{nm4} S. Xue, Y. Zhang, X. Li. (2020). Non-Markovian effects in quantum dynamics: A systematic approach. \textit{Phys. Rev. A}, 101(3), 032110.
\bibitem{nm2} C. Gardiner, P. Zoller. (2004). Quantum noise: a handbook of Markovian and non-Markovian quantum stochastic methods with applications to quantum optics. Springer Science Business Media.
\bibitem{nm5} B. Yue, S. Xue, Y. Pan, M. Jiang. (2023). Quantum approximate optimization algorithm in non-Markovian quantum systems. \textit{Phys. Scr.}, 98(10), 105104.
\bibitem{nm7} H. P. Breuer, E. M. Laine, J. Piilo, B. Vacchini. (2016). Colloquium: Non-Markovian dynamics in open quantum systems. \textit{Rev. Mod. Phys.}, 88(2), 021002.
%\bibitem{nm6} S. Xue, S. Liu. (2019). Modeling non-Markovian dynamics in quantum systems: Physical constraints and optimization techniques. \textit{Quantum Inf. Process.}, 18(12), 344.



%\bibitem{nm8} I. De Vega, D. Alonso. (2017). Dynamics of non-Markovian open quantum systems. \textit{Rev. Mod. Phys.}, 89(1), 015001.

\bibitem{nm9} S. Xue, R. Wu, W. Zhang, J. Zhang, C. Li, T. J. Tarn. (2012). Decoherence suppression via non-Markovian coherent feedback control. \textit{Phys. Rev. A}, 86(5), 052304.

    \bibitem{nm10} S. Xue, M. R. Hush, I. R. Petersen. (2016). Feedback tracking control of non-Markovian quantum systems. \textit{IEEE Trans. Control Syst. Technol.}, 25(5), 1552-1563.



   \bibitem{br} H. P. Breuer, F. Petruccione. (2002). The theory of open quantum systems. OUP Oxford.

 \bibitem{alireza}   C. Sutherland, T. A. Brun, D. A. Lidar. (2018). Non-Markovianity of the post-Markovian master equation. \textit{Phys. Rev. A}, 98(4), 042119.



 \bibitem{qip}  S. Xue, R. Wu, T. J. Tarn, I. R. Petersen, I. R. (2015). Witnessing the boundary between Markovian and non-Markovian quantum dynamics: a Green's function approach. \textit{Quantum Inf. Process.}, 14, 2657-2672.

  \bibitem{weimin zhang}  W. Zhang, P. Lo, H. Xiong, M. Tu, F. Nori. (2012). General non-Markovian dynamics of open quantum systems. \textit{Phys. Rev. Lett.}, 109(17), 170402.

\bibitem{nm11} S. Xue, T. Nguyen, M. R. James, A. Shabani, V. Ugrinovskii, I. R. Petersen. (2020). Modeling for non-Markovian quantum systems. \textit{IEEE Trans. Control Syst. Technol.}, 28(6), 2564-2571.

  \bibitem{nmCDC}  S. Xue, M. R. James, A. Shabani, V. Ugrinovskii, I. R. Petersen. (2015)
    Quantum filter for a class of non-Markovian quantum systems, \textit{54th IEEE Conference on Decision and Control (CDC)}, 7096-7100.

  \bibitem{nmQIP}
  S. Xue, I. R. Petersen. (2016)
Realizing the dynamics of a non-Markovian quantum system by Markovian coupled oscillators: a Green's function-based root locus approach.
\textit{
Quantum Inf. Process.}, 15(2), 1001-1018


  \bibitem{white} V. P. Belavkin. (1995). Quantum filtering of Markov signals with white quantum noise. Quantum communications and measurement. Boston, MA: Springer US, 381-391.

\bibitem{r8} G. Zhang, H. W. Joseph Lee, B. Huang, H. Zhang. (2012). Coherent feedback control of linear quantum optical systems via squeezing and phase shift. \textit{SIAM J. Control Optim.}, 50(4), 2130-2150.

  \bibitem{direct}
    G. Zhang, M. R. James. (2010). Direct and indirect couplings in coherent feedback control of linear quantum systems. \textit{IEEE Trans. Autom. Control}, 56(7), 1535-1550.

\bibitem{ccr}
M. R. James, H. I. Nurdin, and I. R. Petersen. (2008).  $\mathscr{H}_\infty$ control of linear quantum systems, \textit{IEEE Trans. Autom. Control}, 53, (8), pp. 1787-1803.

\bibitem{hur}
J. E. Gough and G. Zhang. (2015). On realization theory of quantum linear systems, \textit{Automatica},
59: 139-151.

\bibitem{robust}
K. Zhou, J. C. Doyle. (1998). Essentials of robust control. Upper Saddle River, NJ: Prentice hall.

\bibitem{H2} S. Gugercin, A. C. Antoulas, and C. Beattie. (2008) $\mathscr{H}_2$ model reduction for large-scale linear dynamical systems, \textit{SIAM J. Control Optim.}, 30(2), pp. 609-638.


\bibitem{sedumi}
Computational Optimization Research at Lehigh (COR@L), Lehigh University, SeDuMi v1.21, (2009), http://sedumi.ie.lehigh.edu/.

\bibitem{yalmip}
J. Lofberg. (2004). ``YALMIP: A toolbox for modeling and optimization in MATLAB," \emph{in 2004 IEEE international conference on robotics and automation}, pp. 284-289.

\end{thebibliography}
\end{document}